\documentclass[aps,preprint,amssymb,superscriptaddress]{revtex4-1}
\usepackage{floatrow}
\newfloatcommand{capbtabbox}{table}[\FBwidth]
\usepackage{graphicx}
\usepackage{dcolumn}
\usepackage{bm}
\usepackage{natbib}
\usepackage{CJK}
\usepackage[utf8]{inputenc}
\usepackage[T1]{fontenc}
\usepackage{mathptmx}
\usepackage{gensymb}
\usepackage{graphicx}
\usepackage{epstopdf}
\usepackage{dcolumn}
\usepackage{bm}
\usepackage{amsmath}
\usepackage{mathtools}
\usepackage{relsize}
\usepackage{multirow}
\usepackage[colorlinks]{hyperref}
\begin{document}


\title{Structure-Adaptive Topology Optimization Framework for Photonic Band Gaps with TM-Polarized Sources}
\author{Aditya Bahulikar}
\affiliation{Department of Electrical Engineering and Computer Science, Syracuse University, Syracuse, NY 13210}
\author{Feng Wang}
\affiliation{Department of Electrical Engineering and Computer Science, Syracuse University, Syracuse, NY 13210}
\author{Mustafa Cenk Gursoy}
\affiliation{Department of Electrical Engineering and Computer Science, Syracuse University, Syracuse, NY 13210}
\author{Rodrick Kuate Defo}
\email{rkuatede@syr.edu}
\affiliation{Department of Electrical Engineering and Computer Science, Syracuse University, Syracuse, NY 13210}

\date{\today}






\begin{abstract}
We present a structure-adaptive topology optimization framework for engineering photonic band gaps with TM-polarized sources based on computation of the photonic density of states with a uniform source substituting for the standard Dirac delta function sources in formalisms analogous to $\Gamma$-point integration and to integration over a full Brillouin zone. We generalize the limiting uniform and Dirac delta function sources to more general collections of sources, such that the union of the sources in a given collection is hyperuniform. The uniform-source approach necessarily leads to the fastest computations. We also demonstrate how our approach can be generalized to the treatment of the frequency-dependent optical response of materials. Finally, we show that we can recover known two-dimensional photonic crystals for the TM polarization. A key advantage of our work is its ability to optimize for a specific midgap frequency and band gap in a structure-adaptive manner. Our work leverages the insight that the determination of the minimum supercell size and the minimum precision to which the frequencies within the photonic band gap must be sampled will lead to the observation of photonic-crystal structures when the $\Gamma$-point formalism for the uniform-source approach is employed. Additionally, our $\Gamma$-point and full Brillouin zone formalisms for the uniform-source approach inherently encourage binarized designs even in gradient descent.
\end{abstract}

\maketitle


\section{Introduction}
Regarding circuits based on semiconductor technology, increased miniaturization is leading to circuits with higher resistance and higher levels of power dissipation~\cite{joannopoulos_photonic_1997,shalf_future_2020}. Photonic systems outperform electronic systems for low-loss, high-capacity transmission of information~\cite{agrawal_fiber_2012,winzer_fiber_2018}, which suggests that fabrication of the analog of a transistor for photonic systems would have the potential to revolutionize the information technology industry~\cite{joannopoulos_photonic_1997}. The difficulty lies in robustly designing structures, known as photonic crystals, with photonic band gaps, as these structures would require feature sizes of less than 1~$\mu$m for their regime of operation~\cite{joannopoulos_photonic_1997,men_robust_2014}. In recent decades, significant advances have been made in the design of nanophotonic structures, including of photonic crystals, as well as in the determination of bounds on the optimal performance of nanophotonic devices~\cite{jensen_topology_2011,molesky_inverse_2018,li_topology_2019,chao_physical_2022,sheverdin_photonic_2020}. Optimizing the distribution of material within a nanophotonic device, as captured by the value of the permittivity $\epsilon$ throughout the design region, was initially accomplished by designing structures by hand or by varying a few parameters to find an optimal solution~\cite{maldovan_diamond_2004,joannopoulos_photonic_2011,fan_design_1994,doosje_photonic_2000,biswas_three_2002,maldovan_photonic_2002,toader_photonic_2003,michielsen_photonic_2003,maldovan_exploring_2003,maldovan_photonic_2005}. Later efforts involved level-set descriptions~\cite{kao_maximizing_2005,burger_inverse_2004,burger_a_2005,he_incorporating_2007} and continuous relaxations of $\epsilon$~\cite{dobson_maximizing_1999,cox_band_2000,shen_large_2002,halkjaer_maximizing_2006,sigmund_geometric_2008,sigmund_systematic_2003,jensen_systematic_2004,watanabe_broadband_2006,liang_formulation_2013}, where every pixel or voxel is allowed to vary continuously within some interval $[\epsilon_{\text{min}},\epsilon_{\text{max}}]$, an approach we adopt in this work though it is not integral to our framework. The optimization algorithms include gradient-based approaches~\cite{cox_band_2000,kao_maximizing_2005,dobson_maximizing_1999}, which we employ in this work for illustrative purposes though employing gradient descent is again not integral to the framework, semidefinite programming~\cite{men_bandgap_2010,men_robust_2014}, and gradient-free techniques~\cite{goh_genetic_2007,yan_photonic_2021,kim_automated_2023}.  Previous approaches to optimizing photonic crystals have involved consideration of the photonic bandstructure and optimization of the gap between consecutive bands in the bandstructure. Such approaches are limited in that they cannot control the size of the band gap or the precise location of the midgap frequency in the optimizations.  

Our approach addresses this shortcoming of previous approaches through its ability to adapt to a given design region structure and optimize for specified target midgap frequencies and band gaps by minimizing the photonic density of states over a frequency window $\Delta \omega$ with central frequency $\omega_0$, where the photonic density of states is computed by employing a uniform source in the place of the standard Dirac delta functions sources in formalisms analogous either to $\Gamma$-point integration or to integration over a full Brillouin zone. We generalize the limiting cases of uniform sources and Dirac delta function sources to more general collections of sources, such that the union of the sources in a given collection is hyperuniform~\cite{torquato_local_2003,siedentop_stealthy_2024,florescu_designer_2009,man_isotropic_2013,yu_engineered_2021,florescu_complete_2009,man_photonic_2013,rechtsman_optimized_2008,wilts_evolutionary_2018,milovsevic_hyperuniform_2019,florescu_optical_2013,klatt_phoamtonic_2019,florescu_effects_2010,klatt_gap_2021,riganti_effective_2024}. Employing the uniform-source approach, we are able to observe paracrystalline structures with local crystalline order, such as found in amorphous silicon~\cite{treacy_the_2012}, as well as crystalline structures. We demonstrate that for a minimum supercell size and a minimum precision to which the frequency window $\Delta\omega$ must be sampled, photonic-crystal structures are observed when the $\Gamma$-point formalism for the uniform-source approach is employed, and our $\Gamma$-point and full Brillouin zone formalisms for the uniform-source approach inherently encourage binarized designs even in gradient descent.  We also show that the frequency-dependent optical response of materials is readily treated under our formalism, unlike in prior approaches. The two-dimensional formulation we present in this work may find applications in investigations of low-dimensional materials~\cite{mattheakis_epsilon_2016,kuate_methods_2021,kuate_strain_2016,shiang_ab_2015}.

\section{Theoretical Approach and Discussion}
{\label{sec:methodology}
\subsection{Deriving the photonic-crystal objective}
In order to create structures with photonic band gaps, which are intervals of frequencies $[\omega_0-\Delta\omega/2,\omega_0+\Delta\omega/2]$ for which no electromagnetic wave can penetrate the structure, we first observe that the existence of a photonic band gap is equivalent to the suppression of the photonic density of states (DOS$(\omega)$) over the frequency interval $[\omega_0-\Delta\omega/2,\omega_0+\Delta\omega/2]$. In order to obtain DOS$(\omega)$, we start with the per-polarization photonic local density of states (LDOS$_j\left(\omega,\mathbf{r}'\right)$), which is related to the power radiated by a point dipole, that is a current $\mathbf{J}_{j,\mathbf{r}'}(\mathbf{r},t) = \hat{e}_je^{-i\omega t}\delta\left(\mathbf{r}-\mathbf{r}'\right)$ where $\hat{e}_j$ is the unit vector in the dipole's direction, $\mathbf{r}'$ is the position of the dipole, $\delta\left(\mathbf{r}-\mathbf{r}'\right)$ is a Dirac delta function that integrates to unity if $\mathbf{r}'$ lies within the region of integration and zero otherwise, and $\omega$ is the frequency of the dipole, in the following manner~\cite{liang_formulation_2013,daguanno_electromagnetic_2004,novotny_principles_2012,inoue_photonic_2004,martin_electromagnetic_1998,taflove_advances_2013},
\begin{equation}
\label{eq:partialLDOS}
\text{LDOS}_j\left(\omega,\mathbf{r}'\right) = -\frac{6}{\pi}\text{Re}\left[\int\mathbf{J}_{j,\mathbf{r}'}^*(\mathbf{r})\cdot\mathbf{E}_{j,\mathbf{r}'}(\omega,\mathbf{r})\text{d}\mathbf{r}\right].
\end{equation}
The electric field appearing in Eq. (\ref{eq:partialLDOS}) is obtained by inverting the equation
\begin{equation}
\label{eq:Maxwell}
\mathcal{M}(\epsilon,\omega)\mathbf{E}_{j,\mathbf{r}'}(\omega,\mathbf{r}) = i\omega\mathbf{J}_{j,\mathbf{r}'}(\mathbf{r})
\end{equation}
where
\begin{equation}
\mathcal{M}(\epsilon,\omega) = \mathbf{\nabla}\times\frac{1}{\mu(\mathbf{r})}\mathbf{\nabla}\times~ -~ \epsilon(\mathbf{r})\omega^2
\end{equation}
is the Maxwell operator with $\mu(\mathbf{r})$ and $\epsilon(\mathbf{r})$ being the permeability and permittivity in the design region, respectively, and 
\begin{equation}
\mathbf{J}_{j,\mathbf{r}'}(\mathbf{r}) = \delta(\mathbf{r}-\mathbf{r}')\hat{e}_j.
\end{equation}
This LDOS$_j\left(\omega,\mathbf{r}'\right)$ is a measure of the local response at a given point in space, which is formulated above as being associated with the electric field from a dipolar current source, and captures the density of electric-field modes at the point in space~\cite{oskooi_electromagnetic_2013}.

Once LDOS$_j\left(\omega,\mathbf{r}'\right)$ has been obtained, the photonic density of states (DOS$(\omega)$), which counts the number of allowed photonic states or modes at a particular frequency, is obtained by summing the LDOS$_j\left(\omega,\mathbf{r}'\right)$ over the relevant polarizations and integrating over the entire crystal~\cite{taflove_advances_2013},
\begin{equation}
\text{DOS}(\omega) = \sum_j \int \text{LDOS}_j(\omega,\mathbf{r}')\text{d}\mathbf{r}'.
\end{equation}
The next step is to apply a finite-bandwidth formalism~\cite{shim_fundamental_2019,liang_formulation_2013} for optimizing DOS$(\omega)$ over the interval $[\omega_0-\Delta\omega/2,\omega_0+\Delta\omega/2]$. Explicitly, we define a complex version of DOS$(\omega)$, which we call DOS$'(\omega)$, and integrate over all frequencies DOS$'(\omega)$ multiplied by a window function that has the form,
\begin{equation}
H_{\omega_0,\Delta\omega,N}(\omega) = \frac{c_N(\frac{\Delta\omega}{2})^{2N-1}}{(\omega-\omega_0)^{2N}+(\frac{\Delta\omega}{2})^{2N}}
\end{equation}
for some normalization constant $c_N$, some central frequency $\omega_0$, some bandwidth which we will also refer to as the band gap $\Delta\omega$, and some integer $N$, which we also define as the number of frequencies sampled within the band gap. This window function approaches a rectangular function as $N\rightarrow\infty$. After the integration of DOS$'(\omega)$, which is explicitly given by 
\begin{equation}
\text{DOS}'(\omega) = -\frac{6}{\pi}\sum_j\int\int\mathbf{J}_{j,\mathbf{r}'}^*(\mathbf{r})\cdot\mathbf{E}_{j,\mathbf{r}'}(\omega,\mathbf{r})\text{d}\mathbf{r}\text{d}\mathbf{r}',
\end{equation}
we extract the real part again~\cite{shim_fundamental_2019,liang_formulation_2013}. Applying Cauchy's residue theorem we obtain the real part of the integrated product of DOS$'(\omega)$ and $H_{\omega_0,\Delta\omega,N}(\omega)$, which we label as $\tilde{\text{DOS}}_N(\omega_0,\Delta\omega)$, as follows 
\begin{align}
&\tilde{\text{DOS}}_N(\omega_0,\Delta\omega) \\&= \text{Re}\left[\int_{-\infty}^\infty\text{DOS}'(\omega)H_{\omega_0,\Delta\omega,N}(\omega)\text{d}\omega\right] \label{eq:omegaint}\\&= \text{Re}\left[\frac{\sum_{n = 0}^{N-1}\left(e^{i(\pi+2\pi n)/(2N)}\right)\text{DOS}'\left(\omega_0-\frac{\Delta\omega}{2}e^{i(\pi+2\pi n)/(2N)}\right)}{\sum_{n=0}^{N-1}e^{i(\pi+2\pi n)/(2N)}}\right]
\\&= \text{Im}\left[\frac{\sum_{n = 0}^{N-1}\left(e^{i(\pi+2\pi n)/(2N)}\right)\text{DOS}'\left(\omega_0-\frac{\Delta\omega}{2}e^{i(\pi+2\pi n)/(2N)}\right)}{\csc\left(\frac{\pi}{2N}\right)}\right].\label{eq:finalresidue}
\end{align}
The procedure is then to minimize the quantity $\tilde{\text{DOS}}_N(\omega_0,\Delta\omega)$. 

We can motivate the application of Cauchy's residue theorem by considering a function with a simple pole at $\omega'$, $\frac{1}{(\omega-\omega')}$, multiplied by a function $f(\omega)$ which is holomorphic in the neighborhood of $\omega'$. Suppose we wish to integrate the product of these functions over a contour $C$ that is a circle of radius $\xi$ centered at $\omega'$ in the plane of complex $\omega$. Since $f(\omega)$ is holomorphic in the neighborhood of $\omega'$, it will have a Taylor expansion in the neighborhood of $\omega'$ given by
\begin{equation}
f(\omega) = \sum_{n=0}^\infty\frac{d^n}{d\omega^n}(f(\omega))|_{\omega =\omega'}\cdot (\omega-\omega')^n.
\end{equation} 
Thus,
\begin{equation}
\int_C\frac{f(\omega)}{(\omega-\omega')}\text{d}\omega = \int_C\sum_{n=0}^\infty \frac{d^n}{d\omega^n}(f(\omega))|_{\omega=\omega'}\cdot (\omega-\omega')^{n-1}\text{d}\omega.
\end{equation}
We note that a point on a circle of radius $\xi$ centered at the origin in the plane of complex $\omega$ can be expressed as $\xi(\cos(t)+i\sin(t))$, where $t$ starts at $0$ and ends at $2\pi$. By Euler's formula, we can therefore write $(\omega-\omega') = \xi e^{it}$ so that $\text{d}\omega = i \xi e^{it}\text{d}t = i(\omega-\omega') \text{d}t$. Thus,
\begin{equation}
\int_C\frac{f(\omega)}{(\omega-\omega')}\text{d}\omega = \int_C\sum_{n=0}^\infty \frac{d^n}{d\omega^n}(f(\omega))|_{\omega=\omega'}\cdot (\omega-\omega')^{n-1}\text{d}\omega = i\int_0^{2\pi}\sum_{n=0}^\infty\frac{d^n}{d\omega^n}(f(\omega))|_{\omega=\omega'}\cdot \xi^n e^{itn}\text{d}t.
\end{equation}
For $n \neq 0$, the integration yields
\begin{equation}
i\int_0^{2\pi}\frac{d^n}{d\omega^n}(f(\omega))|_{\omega=\omega'}\cdot \xi^n e^{itn}\text{d}t = i\frac{d^n}{d\omega^n}(f(\omega))|_{\omega=\omega'}\left.\frac{\xi^n e^{itn}}{in}\right|_0^{2\pi} = i\frac{d^n}{d\omega^n}(f(\omega))|_{\omega=\omega'}\cdot \frac{\xi}{in}^n(1-1) = 0.
\end{equation}
For $n=0$, we obtain $2\pi i f(\omega')$. Thus, 
\begin{equation}
\int_C\frac{f(\omega)}{(\omega-\omega')}\text{d}\omega = 2\pi i f(\omega').
\end{equation}
Given that $\omega'$ and $\xi$ were arbitrary, we can construct any positively oriented simple closed contour $\gamma$ from circular contours of varying radii and can consider any meromorphic function $g(\omega)$ with an arbitrary but finite number of simple poles $\omega_k$ lying within $\gamma$ so that the integral of $g(\omega)$ over $\gamma$ will be given by 
\begin{equation}
\int_\gamma g(\omega)\text{d}\omega = 2\pi i \sum_k\text{Res}(g,\omega_k)
\end{equation}
where 
\begin{equation}
\text{Res}(g,\omega_k) = \lim_{\omega\rightarrow \omega_k}(\omega-\omega_k)g(\omega).\label{eq:residues}
\end{equation}

Using these insights, we can rewrite the integral in Eq. (\ref{eq:omegaint}) as a sum of contributions from the contours $C_1$, $C_2$, and $C_3$ that are depicted in Fig. \ref{fig:contours} for the case of $N = 3$. Given that the density of photonic states scales as $\omega^2$ in three-dimensional free space, for $N \geq 2$ the contribution from $C_3$ will vanish in the limit where we take the radius associated with $C_3$ to infinity. Similarly, the simple pole at the origin associated with the contour $C_2$, whose associated radius we will take to zero, involves the evaluation of $H_{\omega_0,\Delta\omega,N}(0)$~\cite{shim_fundamental_2019,liang_formulation_2013}, which vanishes for sufficiently large $N$ or for sufficiently small $\Delta \omega$ and large $\omega_0$. Given that the closed contour $C_1\cup C_2\cup C_3$ is negatively oriented, 
\begin{align}
\int_{-\infty}^\infty\text{DOS}'(\omega)H_{\omega_0,\Delta\omega,N}(\omega)\text{d}\omega &= \int_{C_1} \text{DOS}'(\omega)H_{\omega_0,\Delta\omega,N}(\omega)d\omega + 0 + 0 \\&=\int_{C_1\cup C_2 \cup C_3} \text{DOS}'(\omega)H_{\omega_0,\Delta\omega,N}(\omega)d\omega \\&= -2\pi i \sum_k\text{Res}(\text{DOS}'H_{\omega_0,\Delta\omega,N},\omega_k).
\end{align}

To find the poles of the function $H_{\omega_0,\Delta\omega,N}(\omega)$, we note that they can be obtained by setting the denominator of $H_{\omega_0,\Delta\omega,N}(\omega)$ equal to zero. In that case, we find that $\omega$ satisfies 
\begin{equation}
\omega = \omega_0 - \left(-\left(\frac{\Delta\omega}{2}\right)^{2N}\right)^{\frac{1}{2N}}.
\end{equation}
We can write $-1 = e^{i\left(\pi+2\pi n\right)}$ where $n$ is an integer less than $2N$ and greater than or equal to zero for uniqueness of the roots. Thus, the roots $\omega_{n+1}$ are given by
\begin{equation}
\omega_{n+1} = \omega_0 - \frac{\Delta \omega}{2}e^{i\left(\pi+2\pi n\right)/(2N)}.
\end{equation}
With the exception of the residue at the origin, the residues associated with the function \\$\text{DOS}'(\omega)H_{\omega_0,\Delta\omega,N}(\omega)$ are given by
\begin{align}
\text{Res}(\text{DOS}'H_{\omega_0,\Delta\omega,N},\omega_{n+1}) &= \lim_{\omega\rightarrow\omega_{n+1}}\left((\omega - \omega_{n+1})\frac{c_N\left(\frac{\Delta\omega}{2}\right)^{2N-1}}{\prod_{m}(\omega-\omega_{m+1})}\text{DOS}'(\omega)\right)\\
&= \lim_{\omega\rightarrow\omega_{n+1}}\left(\frac{c_N\left(\frac{\Delta\omega}{2}\right)^{2N-1}}{\prod_{m\neq n}(\omega-\omega_{m+1})}\text{DOS}'(\omega)\right)\\
&=\frac{c_N\left(\frac{\Delta\omega}{2}\right)^{2N-1}}{\prod_{m\neq n}(\omega_{n+1}-\omega_{m+1})}\text{DOS}'(\omega_{n+1})\\
&= \frac{c_N}{\prod_{m\neq n}(e^{i(\pi+2\pi m)/(2N)}-e^{i(\pi+2\pi n)/(2N)})}\text{DOS}'(\omega_{n+1})\\
&= \frac{c_Ne^{-i(2N-1)(\pi+2\pi n)/(2N)}}{\prod_{m\neq n}(e^{i(2\pi (m-n))/(2N)}-1)}\text{DOS}'(\omega_{n+1})\\
&= \frac{-c_Ne^{i(\pi+2\pi n)/(2N)}}{\prod_{m\neq n}(e^{i(2\pi (m-n))/(2N)}-1)}\text{DOS}'(\omega_{n+1}).
\end{align}
Given that $m$ and $m+2N$ produce equivalent exponentials, without loss of generality, we can relabel $m' = m-n$ and write
\begin{align}
\text{Res}(\text{DOS}'H_{\omega_0,\Delta\omega,N},\omega_{n+1})
&= \frac{-c_Ne^{i(\pi+2\pi n)/(2N)}}{\prod_{m' = 1}^{2N-1}(e^{i(2\pi m')/(2N)}-1)}\text{DOS}'(\omega_{n+1}).\label{eq:residueexpression}
\end{align}
We next observe that the product in the denominator of Eq. (\ref{eq:residueexpression}) is independent of $n$ and since $c_N$ represents the normalization of the function $H_{\omega_0,\Delta\omega,N}(\omega)$, we have from the residue theorem that
\begin{equation}
c_N = \frac{\prod_{m' = 1}^{2N-1}(e^{i(2\pi m')/(2N)}-1)}{-2\pi i\sum_{n=0}^{N-1} \left(-e^{i(\pi+2\pi n)/(2N)}\right)} = \frac{\prod_{m' = 1}^{2N-1}(e^{i(2\pi m')/(2N)}-1)}{2\pi i\sum_{n=0}^{N-1} \left(e^{i(\pi+2\pi n)/(2N)}\right)}.
\end{equation}
From the residue theorem and the expressions for $\text{Res}(H_{\omega_0,\Delta\omega,N}(\omega)\text{DOS}'(\omega),\omega_{n+1})$ and $c_N$, Eq. (\ref{eq:finalresidue}) follows immediately. 

\begin{figure*}[ht!] 
\centering
\includegraphics[width=0.63\textwidth]{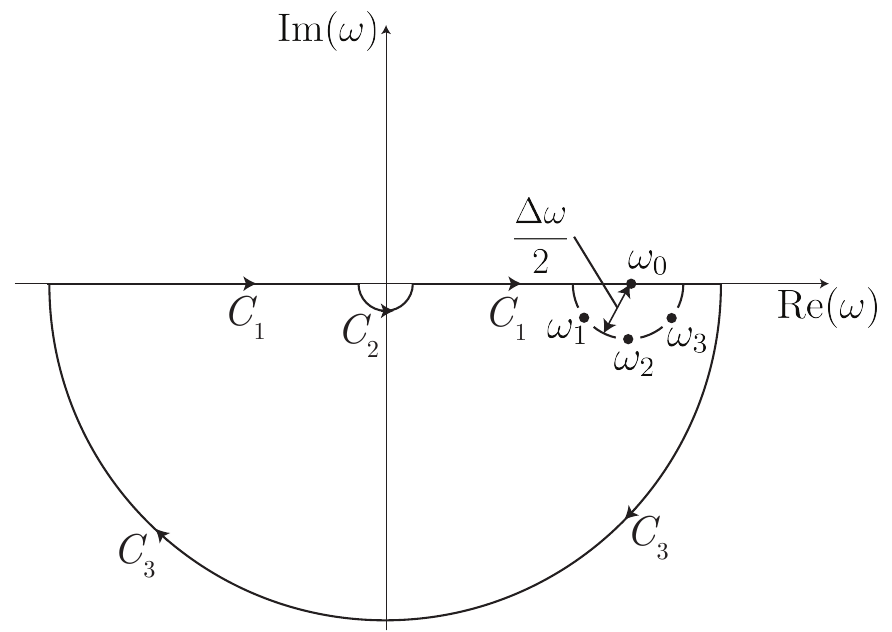}
\caption{Depiction of contours associated with the integral in Eq. (\ref{eq:omegaint}) for the case of $N= 3$.} 
\label{fig:contours}
\end{figure*}

\subsection{Point-dipole computation of the photonic density of states}
\label{sec:pda}
There are at least three ways to approach the problem of computing the $\tilde{\text{DOS}}_N(\omega_0,\Delta\omega)$. The first is computationally intensive. The approach involves inverting Eq. (\ref{eq:Maxwell}) for each frequency  $\omega_{n+1} = \omega_0-\frac{\Delta\omega}{2}e^{i(\pi+2\pi n)/(2N)}$ and for a number of dipoles, $\mathbf{J}_{j,\mathbf{r}'}(\mathbf{r}) = \delta(\mathbf{r}-\mathbf{r}')\hat{e}_j$, equal to the number of pixels or voxels in the design region multiplied by the number of relevant polarizations. There is in principle some inefficiency in such inversions because we really only need the electric field from a given dipole at the location of the dipole. Effectively, to be able to treat all the dipoles needed in the construction of the $\tilde{\text{DOS}}_N(\omega_0,\Delta\omega)$ simultaneously, we would need to be able to operate on a uniform array of dipoles covering the entire design region with the diagonal part of the inverse of the sparse Maxwell operator, that is with the diagonal part of the Green's function, $\mathcal{G} = \mathcal{M}^{-1}$. Since the current vector is composed of point dipoles that integrate to unity, for a single polarization this approach to computing the $\tilde{\text{DOS}}_N(\omega_0,\Delta\omega)$ amounts to computing the trace of $\mathcal{G}$ multiplied by $\frac{6\omega}{\pi}$ and to taking the imaginary part after integrating its product with the window function $H_{\omega_0,\Delta\omega,N}(\omega)$ over all frequencies~\cite{hasan_finite_2018,sprik_optical_1996,busch_photonic_1998,chao_maximum_2022}. Explicitly,
\begin{equation}
\tilde{\text{DOS}}_N(\omega_0,\Delta\omega) = \text{Im}\left[\int_{-\infty}^{\infty}\omega\frac{6}{\pi}\text{Tr}(\mathcal{G})H_{\omega_0,\Delta\omega,N}(\omega)\text{d}\omega\right].
\end{equation}
To our knowledge, there does not exist a sufficiently efficient approach for evaluating the trace of the Green's function for a large sparse Maxwell operator. Indeed, this approach becomes computationally intractable for all but the smallest system sizes, for which the results are not physically meaningful as we will see below. 

\subsection{Uniform-source approach}
\label{sec:usrca}
The second approach involves inverting Eq. (\ref{eq:Maxwell}) once for each frequency and for each relevant polarization, but for uniform currents, $\mathbf{J}_j(\mathbf{r}) = \frac{\hat{e}_j}{V_{pc}}$, covering every pixel or voxel in the design region. The quantity $V_{pc}$ is the volume of the primitive unit cell. Generally, this approach can give reasonable results in the far-field region ($k_{n+1}|\mathbf{r}-\mathbf{r}'| \gg 1$) since for moderate bandwidths the electric field emanating from the location $\mathbf{r}'$ for a given $\omega_{n+1}$ scales approximately as~\cite{jackson_classical_1998} $\mathbf{E} = \omega_0f(|\mathbf{r}-\mathbf{r}'|) e^{ik_{n+1}|\mathbf{r}-\mathbf{r}'|}(\hat{r}\times\hat{p})\times\hat{r}$, where the wavenumber, $k_{n+1}$, is related to the frequency $\omega_{n+1}$, $\hat{r}$ is the unit vector in the direction of $\mathbf{r}-\mathbf{r}'$, and $\hat{p}$ is the unit vector in the direction of the electric dipole moment $\mathbf{p}$. This electric dipole moment $\mathbf{p}$ is obtained as 
\begin{equation}
\mathbf{p} = \int\mathbf{r}'\rho(\mathbf{r}')\text{d}\mathbf{r}',
\end{equation}
where $\rho(\mathbf{r})$ is obtained through
\begin{equation}
i\omega\rho(\mathbf{r}) = \mathbf{\nabla}\cdot \mathbf{J}_{j,\mathbf{r}'}(\mathbf{r}),
\end{equation} 
and where we consider the Kronecker delta contribution, $\mathbf{J}_{j,\mathbf{r}'}(\mathbf{r}) = \delta_{\mathbf{r},\mathbf{r}'}\frac{\hat{e}_j}{V_{pc}}$, to a given uniform current, $\mathbf{J}_j(\mathbf{r}) = \frac{\hat{e}_j}{V_{pc}}$, from each position $\mathbf{r}'$ separately. Summing the electric fields emanating from the location $\mathbf{r}'$ for each of the frequencies $\omega_{n+1}$ leads to the resulting field scaling as $\mathbf{E} = N\omega_0f(|\mathbf{r}-\mathbf{r}'|) \delta_{|\mathbf{r}-\mathbf{r}'|,0}(\hat{r}\times\hat{p})\times\hat{r}$ in the limit where $N\rightarrow\infty$, where $\delta_{|\mathbf{r}-\mathbf{r}'|,0}$ is the Kronecker delta that is equal to 1 when $\mathbf{r} = \mathbf{r}'$ and 0 otherwise and where we have used the relation $\sum_{n=0}^{N-1}e^{ik_{n+1}|\mathbf{r}-\mathbf{r}'|} = N\delta_{|\mathbf{r}-\mathbf{r}'|,0}$ as $N\rightarrow \infty$. Ultimately, the electric field effectively behaves as if we had found the diagonal elements of the inverse of the Maxwell operator and acted on the uniform currents with the operator consisting of those diagonal elements, which is what we desired above.

This uniform-source approach can be motivated by considering again our wave equation with a source
\begin{equation}
\label{eq:Maxwellagain}
\mathcal{M}(\epsilon,\omega)\mathbf{E}_{j,\mathbf{r}'}(\omega,\mathbf{r}) = i\omega\mathbf{J}_{j,\mathbf{r}'}(\mathbf{r})
\end{equation}
and averaging over all $\mathbf{r}'$. In doing so we obtain
\begin{equation}
\label{eq:Maxwellaveraged}
\mathcal{M}(\epsilon,\omega)\frac{1}{N_{pc}V_{pc}}\int\mathbf{E}_{j,\mathbf{r}'}(\omega,\mathbf{r})\text{d}\mathbf{r}' = i\omega \frac{\hat{e}_j}{V_{pc}},
\end{equation}
where we have used the relations for Dirac delta functions shown below in Section \ref{sec:hcs} and where $N_{pc}$ is the number of primitive unit cells in the entire crystal. We can define an effective field 
\begin{equation}
\tilde{\mathbf{E}}_{j}(\omega,\mathbf{r}) \equiv \frac{1}{N_{pc}V_{pc}}\int\mathbf{E}_{j,\mathbf{r}'}(\omega,\mathbf{r})\text{d}\mathbf{r}'
\end{equation}
and it is this single effective field that the uniform-source approach solves for in the same way that Joannopoulos and others~\cite{joannopoulos_photonic_1997,joannopoulos_photonic_2011} treated a single effective field that obeys Bloch's theorem for photonic crystals rather than the forward and backward propagating wave solutions whose sum obeys Bloch's theorem in the formalism proposed by Lord Rayleigh~\cite{rayleigh_on_1887,novotny_principles_2012}. In a similar vein to earlier work~\cite{joannopoulos_photonic_1997,joannopoulos_photonic_2011}, our single effective field must obey Bloch's theorem if we assume that the permittivity in a given primitive unit cell is periodically repeated throughout the crystal so that we can write 
\begin{equation}
\tilde{\mathbf{E}}_{j}(\omega,\mathbf{r}) = \tilde{\mathbf{E}}_{j,\mathbf{k}}(\omega,\mathbf{r})  = \mathbf{u}_{j,\mathbf{k}}(\omega,\mathbf{r})e^{i\mathbf{k}\cdot\mathbf{r}},
\end{equation}
where $\mathbf{k}$ is a wavevector. Our wave equation with a source then becomes
\begin{equation}
\label{eq:Maxwellbloch}
\left(\mathbf{\nabla}_\mathbf{k}\times\frac{1}{\mu(\mathbf{r})}\mathbf{\nabla}_\mathbf{k}\times~ -~ \epsilon(\mathbf{r})\omega^2\right)
\mathbf{u}_{j,\mathbf{k}}(\omega,\mathbf{r}) = i\omega \frac{\hat{e}_j}{V_{pc}}e^{-i\mathbf{k}\cdot\mathbf{r}},
\end{equation}
where 
\begin{equation}
\mathbf{\nabla}_\mathbf{k} = \mathbf{\nabla}+i\mathbf{k}.
\end{equation}
Optimization over an entire Brillouin zone would then proceed by solving Eq. (\ref{eq:Maxwellbloch}) for all $\mathbf{k}$ in the Brillouin zone using a single real-space primitive unit cell. The uniform-source approach for $\mathbf{k} = 0$ can be viewed as analogous to $\Gamma$-point integration.

This approach will succeed when significant variations in $\epsilon(\mathbf{r})$ are present, which we can understand as follows. We first note that in the expression 
\begin{equation}
\text{DOS}'(\omega) = -\frac{6}{\pi}\sum_j\int\int\mathbf{J}_{j,\mathbf{r}'}^*(\mathbf{r})\cdot\mathbf{E}_{j,\mathbf{r}'}(\omega,\mathbf{r})\text{d}\mathbf{r}\text{d}\mathbf{r}',
\end{equation}
we will be able to replace $\mathbf{E}_{j,\mathbf{r}'}(\omega,\mathbf{r})$ with $\tilde{\mathbf{E}}_{j}(\omega,\mathbf{r})$ and $\mathbf{J}_{j,\mathbf{r}'}^*(\mathbf{r})$ with $\frac{\hat{e}_j}{V_{pc}}$ up to a scaling factor as long as the $\mathbf{E}_{j,\mathbf{r}'}(\omega,\mathbf{r})$ are highly localized functions (in the limit where the $\mathbf{E}_{j,\mathbf{r}'}(\omega,\mathbf{r})$ become some factor multiplied by a Kronecker delta, the replacement would become exact). We also note that, from quantum mechanics, the expression
\begin{equation}
-\frac{\hbar^2}{2m}\frac{d^2\psi}{dx^2} = \alpha \delta(x)\psi + E\psi
\end{equation}
where $\psi$ is the wave function of a particle, $m$ is the particle's mass, $E$ is the particle's energy, and $\hbar$ is the reduced Planck constant, will produce bound (localized) states $\psi$ as long as the energy $E$ is less than zero and $\alpha$ is greater than zero. We can rewrite our wave equation for a particular $\mathbf{E}_{j,\mathbf{r}'}$ as 
\begin{align}
&\left(\mathbf{\nabla}\times\frac{1}{\mu(\mathbf{r})}\mathbf{\nabla}\times~ -~ \epsilon(\mathbf{r})\omega^2\right)\mathbf{E}_{j,\mathbf{r}'}= i\omega \delta(\mathbf{r}-\mathbf{r}')\hat{e}_j\\ &\implies -\nabla^2\mathbf{E}_{j,\mathbf{r}'} = i\omega\mu_0 \delta(\mathbf{r}-\mathbf{r}')\hat{e}_j + \epsilon(\mathbf{r})\mu_0\omega^2\mathbf{E}_{j,\mathbf{r}'},
\end{align}
where we have set $\mu(\mathbf{r}) = \mu_0$ and have used the identity $\nabla\times(\nabla\times \mathbf{E}_{j,\mathbf{r}'}) = \nabla(\nabla\cdot\mathbf{E}_{j,\mathbf{r}'})-\nabla^2\mathbf{E}_{j,\mathbf{r}'}$, and the fact that we will be treating sources that are polarized perpendicular to the plane in which we will be taking derivatives so that $\nabla(\nabla\cdot\mathbf{E}_{j,\mathbf{r}'}) = 0$. We can write $\omega = \text{Re}[\omega] + i \text{Im}[\omega]$ so that our equation becomes  
\begin{equation}
-\nabla^2\mathbf{E}_{j,\mathbf{r}'} = (i\text{Re}[\omega]-\text{Im}[\omega])\mu_0 \delta(\mathbf{r}-\mathbf{r}')\hat{e}_j + \epsilon(\mathbf{r})\mu_0((\text{Re}[\omega])^2-(\text{Im}[\omega])^2+2i\text{Re}[\omega]\text{Im}[\omega])\mathbf{E}_{j,\mathbf{r}'}.
\end{equation}
We'll simplify to taking derivatives in one direction, $\hat{e}_x$, and let $j$ to correspond to polarization along an axis perpendicular to the $x$ axis so that we obtain
 \begin{align}
-\frac{d^2}{dx^2}\left(E_{j,x'}\right) &= \left(-\text{Im}[\omega]+i\text{Re}[\omega]\right)\mu_0 \delta(x-x') \\&+ \epsilon(x)\mu_0((\text{Re}[\omega])^2-(\text{Im}[\omega])^2+2i\text{Re}[\omega]\text{Im}[\omega])E_{j,x'}.
\end{align}
Given that $\text{Im}[\omega]\neq 0$ and if we assume that $\epsilon(x)$ is locally constant, we have the form of the Schr\"odinger equation with a Dirac delta function potential that can produce localized states. We can then write our solutions in the form
\begin{equation}
E_{j,x'}(\omega,x) = \left\{\begin{array}{lr} S_{j,x',\omega}e^{-ik\cdot(x-x')}+T_{j,x',\omega}e^{ik\cdot(x-x')}& x-x'< 0\\ U_{j,x',\omega}e^{-ik\cdot(x-x')}+V_{j,x',\omega}e^{ik\cdot(x-x')} &x-x' > 0\end{array}\right.
\end{equation}
where 
\begin{equation}
k= k_{\omega}(x) = \left(\text{Re}[\omega]+i\text{Im}[\omega]\right)\sqrt{\epsilon(x)\mu_0}.
\end{equation}
We impose continuity on the solution so that $S_{j,x',\omega}+T_{j,x',\omega} = U_{j,x',\omega}+V_{j,x',\omega}$ and do not include solutions that diverge as $x\rightarrow\pm\infty$ since such solutions will not conserve energy. Since $\text{Im}[\omega] < 0$, we keep the coefficients $U_{j,x',\omega}$ and $T_{j,x',\omega}$. The derivative of $E_{j,x'}$ will be discontinuous due to the Dirac delta function source and we can solve for the jump in the derivative of $E_{j,x'}$ by integrating our wave equation from $x'-\Delta$ to $x'+\Delta$ and then taking $\Delta$ to zero
\begin{align}
-\left[\frac{d\left(E_{j,x'}\right)}{dx}\right]^{x'+\Delta}_{x'-\Delta} &= \left(-\text{Im}[\omega]+i\text{Re}[\omega]\right)\mu_0 \int^{x'+\Delta}_{x'-\Delta}\delta(x-x')dx\\&+ \int^{x'+\Delta}_{x'-\Delta}\epsilon(x)\mu_0((\text{Re}[\omega])^2-(\text{Im}[\omega])^2+2i\text{Re}[\omega]\text{Im}[\omega])E_{j,x'}dx.
\end{align}
Thus,
\begin{align}
&i\left(U_{j,x',\omega}\cdot k_\omega\left(x\rightarrow (x')^+\right)+T_{j,x',\omega}\cdot k_\omega\left(x\rightarrow (x')^-\right)\right)= \left(-\text{Im}[\omega]+i\text{Re}[\omega]\right)\mu_0,
\end{align}
where $k_\omega\left(x\rightarrow (x')^+\right)$ indicates that $k$ is evaluated as $x$ approaches $x'$ from larger values and $k_\omega\left(x\rightarrow (x')^-\right)$ indicates that $k$ is evaluated as $x$ approaches $x'$ from the smaller values. Since we assumed that $\epsilon(x)$ is locally constant and using the continuity of $E_{j,x'}$, we can write
\begin{equation}
i2U_{j,x',\omega}\cdot k_\omega\left(x'\right) = \left(-\text{Im}[\omega]+i\text{Re}[\omega]\right)\mu_0.
\end{equation}
Thus,
\begin{equation}
\text{Re}[U_{j,x',\omega}]\text{Re}\left[k_\omega\left(x'\right)\right]-\text{Im}[U_{j,x',\omega}]\text{Im}\left[k_\omega\left(x'\right)\right] = \frac{\text{Re}[\omega]\mu_0}{2}
\end{equation}
and
\begin{equation}
-\text{Im}[U_{j,x',\omega}]\text{Re}\left[k_\omega\left(x'\right)\right]-\text{Re}[U_{j,x',\omega}]\text{Im}\left[k_\omega\left(x'\right)\right] = -\frac{\text{Im}[\omega]\mu_0}{2}.
\end{equation}
The equations above set the real and imaginary parts of $U_{j,x',\omega}$ to
\begin{equation}
\text{Re}[U_{j,x',\omega}] = \frac{\mu_0}{2}\frac{\text{Re}[\omega]\text{Re}\left[k_\omega\left(x'\right)\right] + \text{Im}[\omega]\text{Im}\left[k_\omega\left(x'\right)\right]}{\left|k_\omega\left(x'\right)\right|^2} = \frac{1}{2}\sqrt{\frac{\mu_0}{\epsilon(x')}}\label{eq:ReU}
\end{equation}
and 
\begin{equation}
\text{Im}[U_{j,x',\omega}] = -\frac{\mu_0}{2}\frac{\text{Re}[\omega]\text{Im}\left[k_\omega\left(x'\right)\right] - \text{Im}[\omega]\text{Re}\left[k_\omega\left(x'\right)\right]}{\left|k_\omega\left(x'\right)\right|^2} = 0.
\end{equation}
Our optimization attempts to minimize a quantity proportional to $-\text{Re}[E_{j,x'}]\delta(x-x')$ integrated over all $x$ and $x'$, so we need to maximize $\text{Re}[U_{j,x',\omega}]$. Thus, Eq. (\ref{eq:ReU}) appears to suggest that we should minimize $\epsilon(x')$, which is consistent with the fact that electric field modes tend to concentrate their density in high-$\epsilon$ regions~\cite{joannopoulos_photonic_2011} so that by minimizing the number of high-$\epsilon$ regions the density of electric field modes will also decrease. Conservation of energy, however, requires that $\text{Re}[U_{j,x',\omega}]$ be minimized for some $x'$. Also, in order for $E_{j,x'}$ to be highly localized, increasing $\left|\text{Im}\left[k_\omega\left(x'\right)\right]\right|$ is desirable. A compromise is to have significant variations in $\epsilon(x)$ so that if $\epsilon(x')$ is small for some $x'$, for some other $x'$ we can have large $\epsilon(x')$ and $\left|\text{Im}\left[k_\omega\left(x'\right)\right]\right|$. Therefore, this approach encourages binarization and is not obstructed by it.

To obtain the electric fields for TM polarization, we treat the structure to optimize as varying in the $xy$ plane with a single uniform current, $\mathbf{J}_z(\mathbf{r}) = \frac{\hat{e}_z}{V_{pc}}$, and use the ceviche code for the scalar electric field~\cite{hughes_forward-mode_2019} with the NLopt package~\cite{johnson_nlopt_2007} employing the Method of Moving Asymptotes (MMA) algorithm~\cite{svanberg_class_2002}. We note that the ceviche code has the time convention $+i\omega t$~\cite{hughes_forward-mode_2019}. The structure and source are understood to be uniform in the $\hat{e}_z$ direction, which we neglect in the integrations. We also note that in our calculations, $V_{pc} = a^2 = (1)^2 = 1$, where $a = 1$ is the lattice constant of the primitive square cell.

\subsection{Hyperuniform collections of sources}
\label{sec:hcs}
The third approach involves combining the point-dipole approach and the uniform-source approach. Explicitly, we consider sources $\mathbf{J}_{j,\mathbf{r}'}(\mathbf{r}) = A \sum_\mathbf{q}\frac{\hat{e}_j}{N_{pc}V_{pc}}e^{i\mathbf{q}\cdot(\mathbf{r}-\mathbf{r}')}$, where the $\mathbf{q}$ are chosen from among some set of vectors, the positions $\mathbf{r}'$ are chosen from among the set of all the positions in the design region that do not create equivalent $\mathbf{J}_{j,\mathbf{r}'}(\mathbf{r})$, and $A$ is a scale factor which is not unity in cases where the number of vectors $\mathbf{q}$ is less than the total number of vectors in the space that is reciprocal to the real-space design region. This form for the sources captures evenly sampling from sources with periodicities specified by the $\mathbf{q}$ vectors, where each vector $\mathbf{q}$ can be understood to represent a wavevector. Again, Eq. (\ref{eq:Maxwell}) must be inverted for all relevant polarizations. The covering of the real-space design region provided by the collection of sources associated with the various positions $\mathbf{r}'$ that do not create equivalent $\mathbf{J}_{j,\mathbf{r}'}(\mathbf{r})$ is hyperuniform by construction and not necessarily the individual sources themselves. Here, hyperuniformity is satisfied by periodic systems and is described by the following constraint on the number variance of points where material (and not vacuum) exists in subdomains $\Omega$ of the entire design region~\cite{torquato_local_2003},
\begin{equation}
\left<N_\Omega^2\right>-\left<N_\Omega\right>^2 \sim R^{d-1},
\end{equation}
where $R$ is the radius of the domain $\Omega$, $d$ is the dimensionality of the system, $N_\Omega$ is the number of points where material exists in the domain $\Omega$, and $\left<...\right>$ denotes an average over all domains $\Omega$ having the same radius $R$. For the point-dipole approach, the vectors $\mathbf{q}$ cover all of the space that is reciprocal to the real-space design region ($A = 1$ is employed). For the uniform-source approach, there is a single vector $\mathbf{q} = 0$ ($A = N_{pc}$ is employed), which can be equivalent to $\Gamma$-point integration or to integration over the full Brillouin zone of reciprocal space depending on the number of vectors $\mathbf{k}$ described in Section \ref{sec:usrca} that are used. Thus, the point-dipole approach and the uniform-source approach are limiting cases of these more general sources.

In deriving the form for the more general hyperuniform collections of sources above we have used the relations~\cite{kaxiras_quantum_2019}
\begin{equation}
\delta(\mathbf{r}-\mathbf{r}') = \frac{1}{(2\pi)^3}\int e^{i(\mathbf{r}-\mathbf{r}')\cdot\mathbf{k}}\text{d}\mathbf{k},
\end{equation}
\begin{equation}
\delta(\mathbf{k}-\mathbf{k}') = \frac{1}{(2\pi)^3}\int e^{-i(\mathbf{k}-\mathbf{k}')\cdot\mathbf{r}}\text{d}\mathbf{r},
\end{equation}
and
\begin{equation}
N_{pc}\text{d}\mathbf{k} = \frac{(2\pi)^3}{V_{pc}},
\end{equation}
where the integrals are over all of reciprocal space and real space, respectively. If we define $N_{\text{BZ}}$ as the number of Brillouin zones in all of reciprocal space and note that Dirac delta functions integrate to unity, we obtain
\begin{equation}
N_{\text{BZ}}\text{d}\mathbf{r} = V_{pc}.
\end{equation}
Just as $N_{pc}$ is also equal to the number of reciprocal-space points in a Brillouin zone, we can identify $N_{\text{BZ}}$ with the number of real-space points in a primitive unit cell.
From these relations, the Dirac delta functions in their discrete and continuous representations must obey the following equivalences,
\begin{equation}
\text{continuous:}~\delta(\mathbf{r}-\mathbf{r}') \leftrightarrow \text{discrete:}~\frac{N_{\text{BZ}}}{V_{pc}}\delta_{\mathbf{r}\mathbf{r}'}
\end{equation}
and
\begin{equation}
\quad~\text{continuous:}~\delta(\mathbf{k}-\mathbf{k}') \leftrightarrow \text{discrete:}~\frac{N_{pc}V_{pc}}{(2\pi)^3}\delta_{\mathbf{k}\mathbf{k}'},
\end{equation}
which are consistent with the expression for the more general hyperuniform collections of sources. 

In the limit of a sufficiently large supercell our uniform-source approach is symmetry agnostic if boundary conditions could be neglected since a uniform source does not exclude the symmetry of any space group. To discover crystals having symmetries that are not enforced by the underlying symmetry of the design region while obeying boundary conditions, a potential approach is to increase the quantity $\omega_0$, which leverages the scale invariance of Maxwell's equations~\cite{joannopoulos_photonic_2011} to compress the crystal in order to reduce the effect of the boundary. 

\subsection{Frequency-dependent optical response}
To treat frequency-dependent optical response when $\epsilon(\mathbf{r}) = \epsilon(\omega,\mathbf{r})$ varies continuously within some interval $[\epsilon_\text{min},\epsilon_\text{max}]$, one would apply the transformation \begin{equation}
\left[\epsilon_\text{min},\epsilon_\text{max}\right]\rightarrow\left[\frac{\epsilon_{\text{min},r}(\omega_{n+1})}{\epsilon_{\text{min},r}(\omega_{m+1})}\epsilon_\text{min},\frac{\epsilon_{\text{max},r}(\omega_{n+1})}{\epsilon_{\text{max},r}(\omega_{m+1})}\epsilon_\text{max}\right]  
\end{equation} 
at each frequency $\omega_{n+1}$, where $\epsilon_{\text{min},r}(\omega_{n+1})$ is the relative permittivity evaluated at the frequency $\omega_{n+1}$ for the material with the smaller permittivity at some reference frequency and $\epsilon_{\text{max},r}(\omega_{n+1})$ is the relative permittivity evaluated at the frequency $\omega_{n+1}$ for the material with the larger permittivity at some reference frequency. We use a reference frequency $\omega_{m+1}$ to define the permittivities $\epsilon_\text{min}$ and $\epsilon_\text{max}$ as $\epsilon_\text{min} = \epsilon_{\text{min},r}(\omega_{m+1})\epsilon_0$ and $\epsilon_\text{max} = \epsilon_{\text{max},r}(\omega_{m+1})\epsilon_0$. Effectively, the treatment of frequency-dependent optical response for a given material amounts to a rescaling of the permittivity bounds by a scale factor that depends on the $\omega_{n+1}$, but not on $\mathbf{r}$, when evaluating the DOS$'$ at a given frequency $\omega_{n+1}$.

\section{Results} \label{sec:results}
\subsection{Photonic-crystal behavior as a function of system size}\label{sec:systemsize}
We now present results for two-dimensional structures supporting photonic band gaps for the transverse magnetic (TM) polarization that were obtained by suppressing the photonic density of states of the structures over a frequency window $\Delta\omega$ with a central frequency $\omega_0$, where the photonic density of states is computed using the uniform-source approach with $\mathbf{k} = 0$. For all TM calculations in this section, we assigned the values $\omega_0 = 0.4\cdot2\pi c/a$, $\Delta\omega = \omega_0/10$, and allowed the permittivity $\epsilon(\mathbf{r})$ to vary between $\epsilon_0$ and $8.9\epsilon_0$ for comparison with the structures found in Ref.~\cite{joannopoulos_photonic_2011}. In all of the calculations presented in this work, we have set $\mu(\mathbf{r}) = \mu_0$ for the permeability. Additionally, all designs in this section were obtained from the optimization of design regions within which each grid point was randomly initialized with a permittivity value between $\epsilon_0$ and $8.9\epsilon_0$. The quantities $c$, $\mu_0$, and $\epsilon_0$ are the speed of light, the permeability, and the permittivity for free space, respectively, and we set $a=1$ since we can treat lengths without dimensions given the scale invariance of Maxwell's equations~\cite{joannopoulos_photonic_2011}. Photonic-crystal structures as a function of the system size and the grid-point resolution (gpr), which is the number of grid points used to resolve a length $a = 1$, are shown for the TM polarization in Fig. \ref{fig:TM_sizes}. By performing a Fourier transform of the structure that was optimized over 5000 iterations for a $10\times10$ design region, a gpr of 100, and $N = 10$, where $N$ is the number of frequencies sampled within the band gap, we can extract the location of the four highest maxima of the absolute value of the transform that occur at non-trivial positions. We find that these maxima are located at $\mathbf{k}_1 = -8\cdot 2\pi/(10a)\hat{e}_x$, $\mathbf{k}_2 = 8\cdot 2\pi/(10a)\hat{e}_x$, $\mathbf{k}_3 = 8\cdot 2\pi/(10a)\hat{e}_y$, and $\mathbf{k}_4 = -8\cdot 2\pi/(10a)\hat{e}_y$. Thus, the structure has the underlying symmetry of the square supercell lattice with a new lattice constant $a' = \frac{10}{8}a$.

\begin{figure*}[ht!] 
\centering
\includegraphics[width=0.8\textwidth]{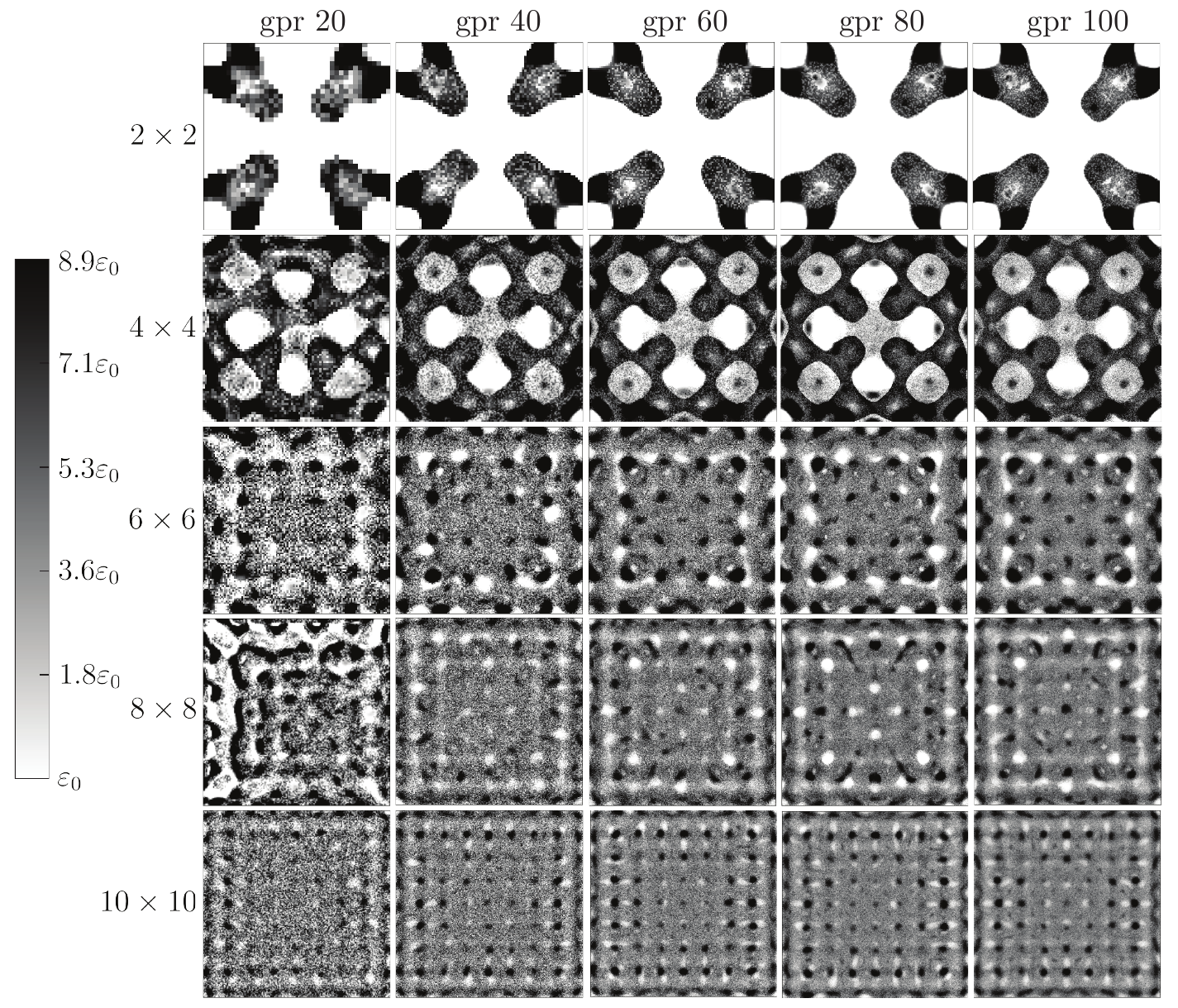}
\caption{Designs optimized for 5000 iterations and $N=10$ with TM polarization as a function of the size of the design region and the number of grid points (gpr) used to resolve a length $a = 1$. The color bar indicates the permittivity $\epsilon$ in the various parts of the design region.} 
\label{fig:TM_sizes}
\end{figure*}

\begin{figure*}[ht!] 
\centering
\includegraphics[width=0.8\textwidth]{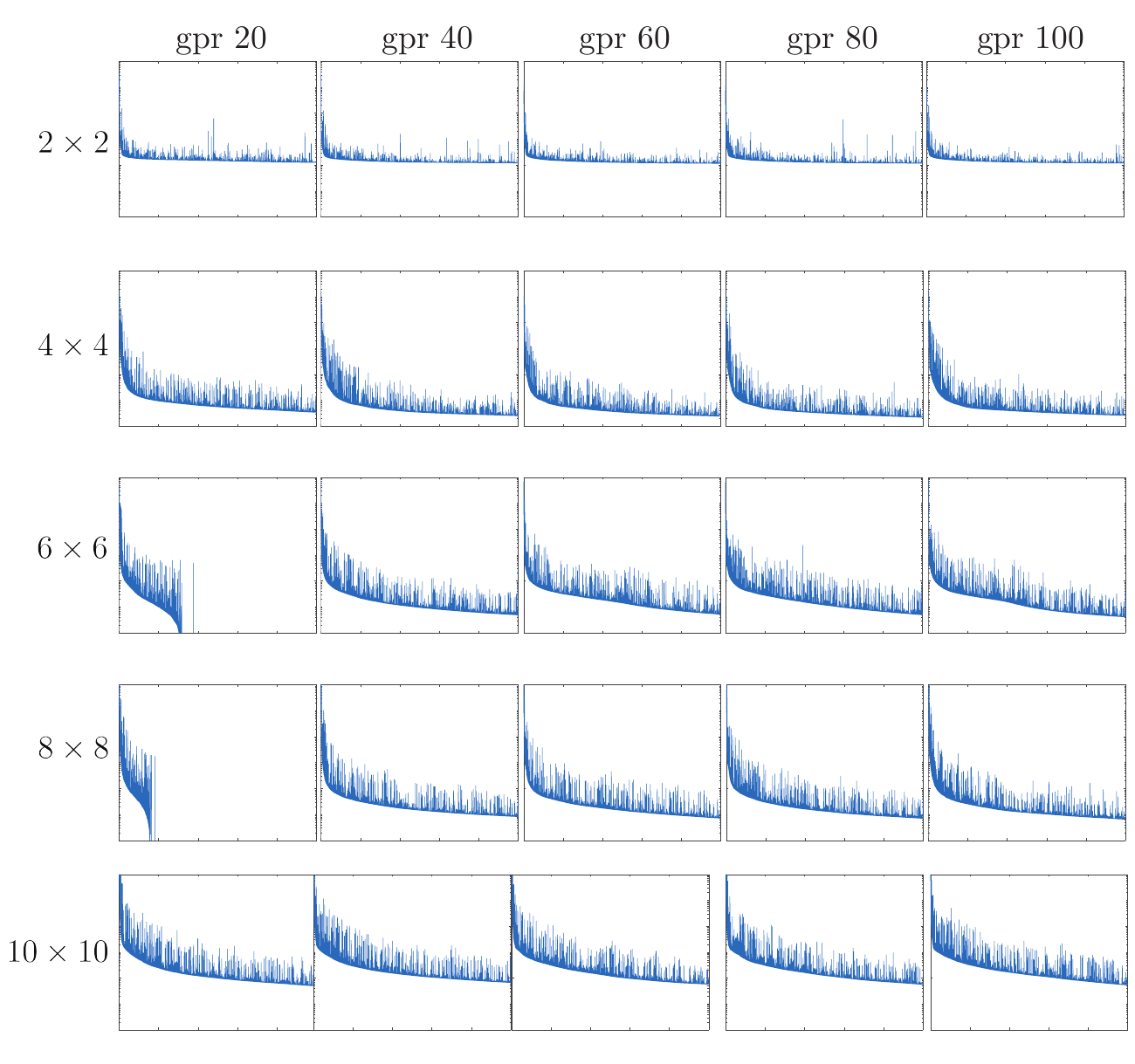}
\caption{Convergence of $\tilde{\text{DOS}}_{10}(\omega_0,\Delta\omega)/\tilde{\text{DOS}}^{\text{vac}}_{10}(\omega_0,\Delta\omega)$ as a function of the number of iterations for various design region sizes and gpr values. The quantities $\tilde{\text{DOS}}^{\text{vac}}_{10}(\omega_0,\Delta\omega)$ are the integrals of the product of $\text{DOS}'(\omega)$ and the window function $H_{\omega_0,\Delta\omega,10}(\omega)$, neglecting the pole at $\omega = 0$, when $\epsilon(\mathbf{r}) = \epsilon_0$ throughout the design region, and are computed for the respective design region sizes and gpr values. The abscissae of all plots are shown for the interval from 0 to 5000 iterations on a linear scale and all ordinate axes are shown for the interval from $10^{-6}$ to 1 on a logarithmic scale.} 
\label{fig:TM_sizes_DOS}
\end{figure*}

The trend toward known photonic-crystal structures as a function of system size can be understood by observing that the Bragg length, which dictates the characteristic length for attenuation of light illuminating a photonic crystal, varies as~\cite{hasan_finite_2018,neve-oz_bragg_2004}
 \begin{equation}
L_B \approx \frac{d}{\ln\left(\frac{1+\sin\left(\frac{\pi\Delta\omega}{4\omega_0}\right)}{1-\sin\left(\frac{\pi\Delta\omega}{4\omega_0}\right)}\right)}\approx \frac{2d}{\pi}\frac{\omega_0}{\Delta\omega},
\end{equation}
where $\omega_0$ is the central frequency, $\Delta\omega$ is the band gap or frequency window, as above, and $d$ is the smallest distance for a set of crystal planes. Thus, given $d = a' = \frac{10}{8}a =  \frac{10}{8}\cdot 1 =  \frac{10}{8}$ and $\frac{\omega_0}{\Delta\omega} = \frac{1}{0.1}$, we expect a minimum crystal linear dimension of $L \gtrsim L_B\approx 8$ in order to observe photonic crystal structures, in good agreement with the results in Fig. \ref{fig:TM_sizes}. Convergence of the quantities $\tilde{\text{DOS}}_{10}(\omega_0,\Delta\omega)/\tilde{\text{DOS}}^{\text{vac}}_{10}(\omega_0,\Delta\omega)$, where the $\tilde{\text{DOS}}^{\text{vac}}_{10}(\omega_0,\Delta\omega)$ are the integrals of the product of $\text{DOS}'(\omega)$ and the window function $H_{\omega_0,\Delta\omega,10}(\omega)$, neglecting the pole at $\omega = 0$, when $\epsilon(\mathbf{r}) = \epsilon_0$ throughout the design region (the $\tilde{\text{DOS}}^{\text{vac}}_{10}(\omega_0,\Delta\omega)$ are computed for the respective design region sizes and gpr values), is shown in Fig. \ref{fig:TM_sizes_DOS}. The results for a gpr of 20 with $6\times6$ as well as $8\times8$ design region sizes encountered numerical instabilities in their convergence and exhibited negative $\tilde{\text{DOS}}_{10}(\omega_0,\Delta\omega)/\tilde{\text{DOS}}^{\text{vac}}_{10}(\omega_0,\Delta\omega)$ values. These instabilities were likely due to the closeness of the design region linear dimensions and the transition point suggested by the Bragg length of  $L_B\approx 8$, and the insufficient resolution to properly capture the transition. The $2\times2$ design region size suppresses $\tilde{\text{DOS}}_{10}(\omega_0,\Delta\omega)/\tilde{\text{DOS}}^{\text{vac}}_{10}(\omega_0,\Delta\omega)$ the least, as expected. An advantage of our approach, as detailed in Ref.~\cite{liang_formulation_2013}, is that it captures the integral of the product of the photonic density of states and a window function very accurately if $N$ is large enough or $\Delta\omega/\omega_0$ is small enough that we can neglect the pole at $\omega = 0$ and is faster than numerically integrating the photonic density of states without a window function over the interval $[\omega_0-\Delta\omega/2,\omega_0+\Delta\omega/2]$ (approximately 9 seconds per iteration on one core for 11 grid points per linear dimension in a $1\times1$ cell with $N = 10$ for full Brillouin zone integration in the uniform-source approach compared to approximately 27 minutes on one core for 11 grid points per linear dimension in a $1\times1$ cell for the SciPy integrate.quad function in the place of the summation over complex frequencies). 

\subsection{Photonic crystal behavior as a function of the number of frequencies, $N$, sampled within the band gap}
We next examine the photonic-crystal behavior as a function of $N$ for TM polarization. As above, all designs in this section were obtained from the optimization of design regions within which each grid point was randomly initialized with a permittivity value between $\epsilon_0$ and $8.9\epsilon_0$.  As shown in Fig. \ref{fig:TM_N}, we find that photonic-crystal behavior is not captured for all $N$. The location of the crossover point to the final TM photonic-crystal structure can be understood by observing that each sampled frequency $\omega_{n+1}$, within the band gap, is in correspondence with a wavenumber $k_{n+1} = \frac{\omega_{n+1}}{v}$, where $v$ is the speed of light through the given medium. These wavenumbers $k_{n+1}$ must be able to capture the length scales of all structures in the photonic crystal. In particular, they must be able to capture the length scale corresponding to the periodicity of the photonic crystal. We treat time-harmonic fields so that we can use the rules for addition of sinusoidal functions to deduce that the smallest difference between these $k_{n+1}$ must be equal to the smallest wavenumber at which a peak is observed in the Fourier transform of the optimized structure. This argument follows from the fact that the function resulting from the sum of sinusoidal functions is a product of an envelope function with frequency equal to the smaller of the difference between or the sum of the frequencies of the original sinusoidal functions and a modulated sinusoidal function with frequency equal to the larger of the difference between or the sum of the frequencies of the original sinusoidal functions. Given a smallest wavenumber $k_{\text{min}} = 8\cdot 2\pi/(10a)$, we find a minimum $N$ given by $N_{\text{min}}\approx \text{gpr}\cdot\frac{\Delta\omega}{vk_\text{min}} \approx5$ for a gpr of 100, $\omega_0 = 0.4\cdot2\pi c/a$, $\Delta\omega = \omega_0/10$, $\frac{\epsilon(\mathbf{r})}{\epsilon_0} \in \left[1,8.9\right]$, and using $v\sim c$, in good agreement with the results shown in Fig. \ref{fig:TM_N}. Physically, the resolution of the grid of an optimized photonic crystal as well as the wavelength associated with the periodicity of the photonic crystal dictate how far apart frequencies must be before they are no longer considered to be identical by the given photonic crystal. The minimum number of frequencies that the photonic crystal must be able to suppress is precisely equal to the maximum number of numerically non-equivalent frequencies. In Fig. \ref{fig:TM_N_DOS}, we show convergence of the quantities $\tilde{\text{DOS}}_{N}(\omega_0,\Delta\omega)/\tilde{\text{DOS}}^{\text{vac}}_{N}(\omega_0,\Delta\omega)$, where the $\tilde{\text{DOS}}^{\text{vac}}_{N}(\omega_0,\Delta\omega)$ are the integrals of the product of $\text{DOS}'(\omega)$ and the window function $H_{\omega_0,\Delta\omega,N}(\omega)$, neglecting the pole at $\omega = 0$, when $\epsilon(\mathbf{r}) = \epsilon_0$ throughout the design region, and are computed for a gpr of 100 and a $10\times10$ design region. Again, $\tilde{\text{DOS}}_{N}(\omega_0,\Delta\omega)$ and $\tilde{\text{DOS}}^{\text{vac}}_{N}(\omega_0,\Delta\omega)$ are computed in the uniform-source approach with $\mathbf{k} = 0$. We observe that the suppression of $\tilde{\text{DOS}}_{N}(\omega_0,\Delta\omega)/\tilde{\text{DOS}}^{\text{vac}}_{N}(\omega_0,\Delta\omega)$ steadily improves with $N$.

\begin{figure*}[ht!] 
\centering
\includegraphics[width=0.8\textwidth]{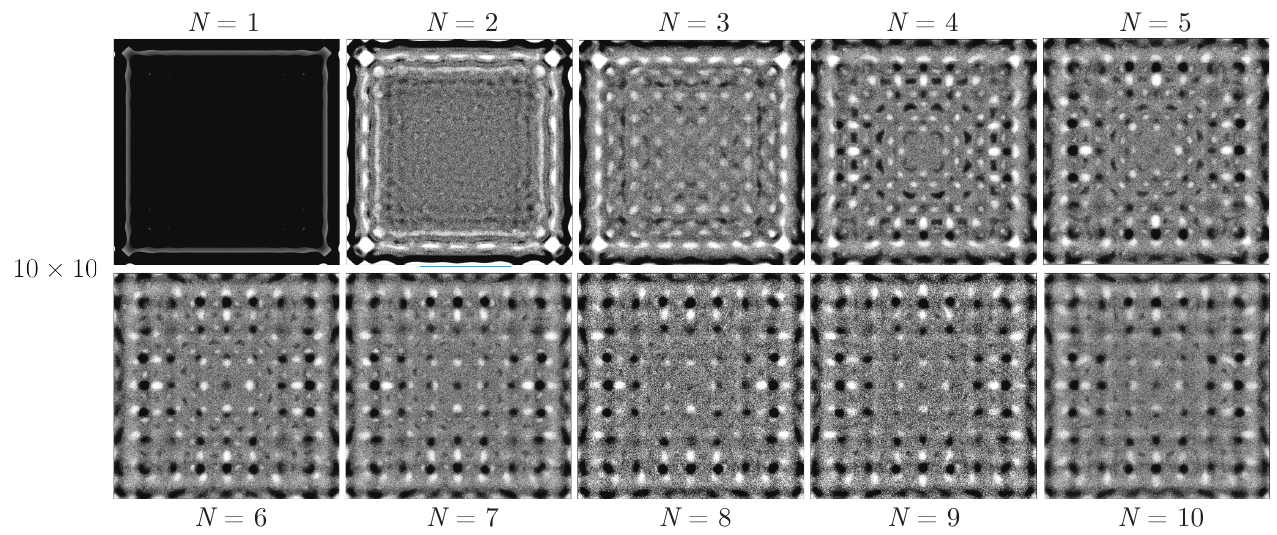}
\caption{Designs optimized for 5000 iterations, a gpr of 100, and a $10\times10$ design region as a function of the number of frequencies sampled within the band gap, $N$. TM polarization was used. Material permittivity values are as in Fig. \ref{fig:TM_sizes}. We include $N=1$ for completeness, though the case of $N=1$ is not physically meaningful due to the fact that the integral in Eq. (\ref{eq:omegaint}) formally diverges for $N=1$ given the scaling of the photonic density of states for two-dimensional free space.} 
\label{fig:TM_N}
\end{figure*}

\begin{figure*}[ht!] 
\centering
\includegraphics[width=0.8\textwidth]{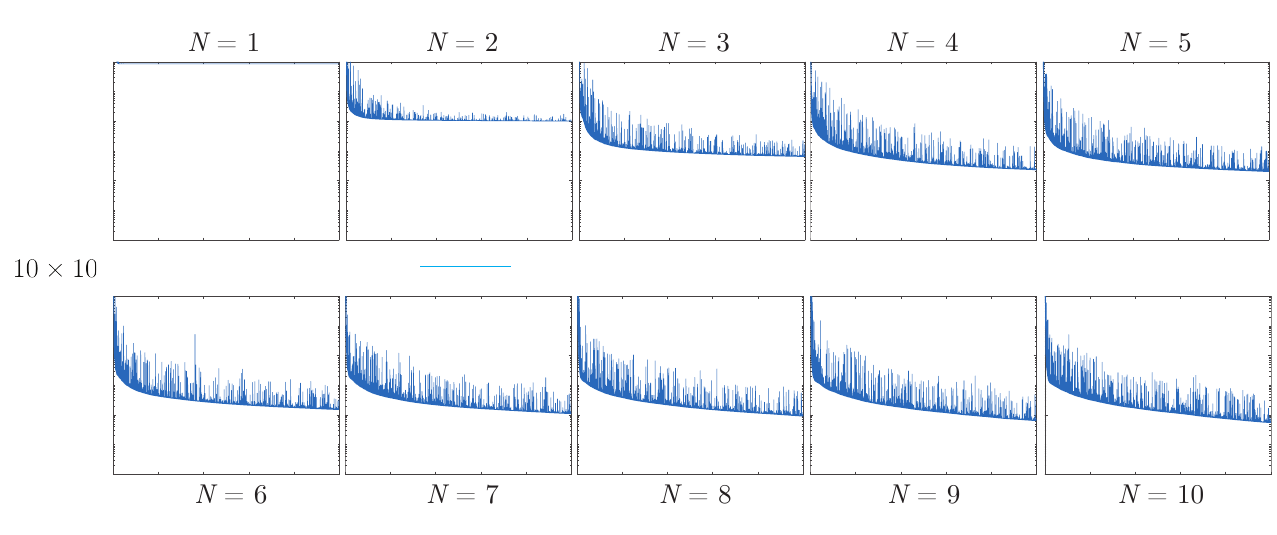}
\caption{Convergence of $\tilde{\text{DOS}}_{N}(\omega_0,\Delta\omega)/\tilde{\text{DOS}}^{\text{vac}}_{N}(\omega_0,\Delta\omega)$ as a function of the number of iterations for various $N$, a gpr of 100,  and a $10\times10$ design region. The quantities $\tilde{\text{DOS}}^{\text{vac}}_{N}(\omega_0,\Delta\omega)$ are the integrals of the product of $\text{DOS}'(\omega)$ and the window function $H_{\omega_0,\Delta\omega,N}(\omega)$, neglecting the pole at $\omega = 0$, when $\epsilon(\mathbf{r}) = \epsilon_0$ throughout the design region, and are also computed for a gpr of 100 and a $10\times10$ design region. The abscissae of all plots are shown for the interval from 0 to 5000 iterations on a linear scale and all ordinate axes are shown for the interval from $10^{-6}$ to 1 on a logarithmic scale.} 
\label{fig:TM_N_DOS}
\end{figure*}

\subsection{Demonstration of the structure-adaptive nature of our approach}
As alluded to above, boundary conditions affect the ability to observe symmetry-agnostic structures. As an illustration, we can consider a dielectric contrast of 3.1, where 6-fold symmetric structures are preferred over 4-fold symmetric structures given a desired band gap of $\Delta\omega = \omega_0/10$~\cite{rechtsman_method_2009}. We therefore assign $\epsilon = 3.1\epsilon_0$ as the maximum value of the permittivity and $\epsilon = \epsilon_0$ as the minimum value of the permittivity and initialize design regions such that each grid point is assigned with a randomly chosen permittivity value between $\epsilon_0$ and $3.1\epsilon_0$ for optimizations with $N = 10$ and a $10\times10$ design region with various $\omega_0$ in order to leverage the scale invariance of Maxwell's equations to compress the crystal in order to reduce the effect of the boundary. We next subdivide the optimized supercells into 10 cells in each dimension (100 cells in total) and compute the location of the four highest maxima of the absolute value of the Fourier transform that occur at non-trivial positions in each of the cells. We find that the design regions partition into subdomains where primitive reciprocal lattice vectors are generally either oriented along the $x$ and $y$ directions or at 45$^\circ$ to the $x$ and $y$ directions (see Fig. \ref{fig:varyomega}). Convergence behavior of $\tilde{\text{DOS}}_{10}(\omega_0,\Delta\omega)/\tilde{\text{DOS}}^{\text{vac}}_{N}(\omega_0,\Delta\omega)$ can be found in Fig. \ref{fig:varyomegaconv}. We typically find on the order of one region per supercell where 6-fold symmetry is obeyed to within an angle of 5$^\circ$. Ultimately, the boundary confines the structure to exhibit 4-fold symmetry and, since 6-fold symmetry is desired, the structure compensates by creating different domains of 4-fold symmetry to minimize $\tilde{\text{DOS}}_{10}(\omega_0,\Delta\omega)/\tilde{\text{DOS}}^{\text{vac}}_{10}(\omega_0,\Delta\omega)$. Thus, our approach is able to capture polycrystalline systems. Similarly, for a maximum value of the permittivity of $\epsilon = 8.9\epsilon_0$ and a minimum value of $\epsilon = \epsilon_0$ using the random initialization described in Section \ref{sec:systemsize}, we generally find 4-fold symmetric subdomains again with primitive reciprocal lattice vectors that are generally either oriented along the $x$ and $y$ directions or at 45$^\circ$ to the $x$ and $y$ directions (see Fig. \ref{fig:varyomega8d9}). In the case of $\frac{\epsilon(\mathbf{r})}{\epsilon_0} \in [1,8.9]$ , we again find on the order of one subdomain that is predominantly described by 6-fold symmetry per supercell. The fact that it is easier to open a gap for larger dielectric contrast (see the suppression of $\tilde{\text{DOS}}_{10}(\omega_0,\Delta\omega)/\tilde{\text{DOS}}^{\text{vac}}_{10}(\omega_0,\Delta\omega)$ in Figs. \ref{fig:varyomegaconv} and \ref{fig:varyomega8d9conv}) is also consistent with known results~\cite{joannopoulos_photonic_2011,rechtsman_method_2009}.

\begin{figure*}[ht!] 
\centering
\includegraphics[width=0.99\textwidth]{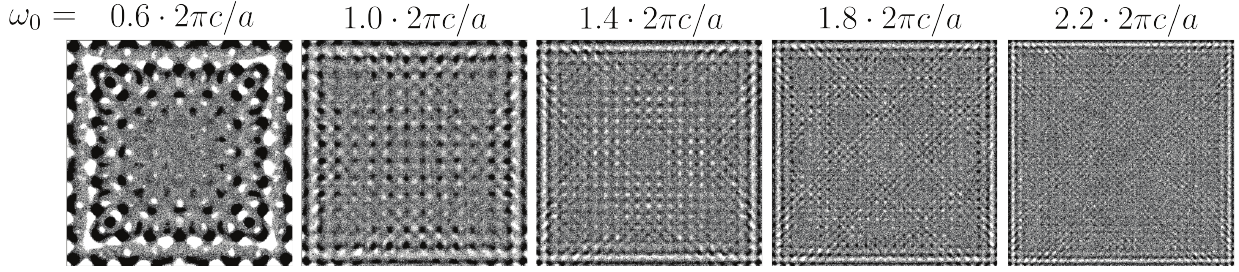}
\caption{Designs optimized for 2500 iterations, a gpr of 100, and a $10\times10$ design region with $N = 10$ as a function of the midgap frequency $\omega_0$ with $\Delta\omega = \omega_0/10$. TM polarization was used. Material permittivity values are in the range $\frac{\epsilon(\mathbf{r})}{\epsilon_0} \in [1,3.1]$ with darker color representing larger permittivity values and lighter color representing smaller permittivity values (the extremal permittivity values have the same coloring as the extremal permittivity values in Fig. \ref{fig:TM_sizes} and the remaining permittivity values are assigned colors according to the same scaling as in Fig. \ref{fig:TM_sizes}).}
\label{fig:varyomega}
\end{figure*}

\begin{figure*}[ht!] 
\centering
\includegraphics[width=0.99\textwidth]{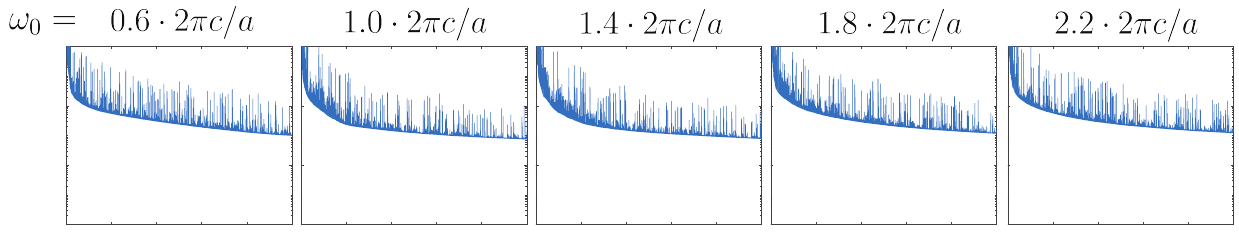}
\caption{Convergence of $\tilde{\text{DOS}}_{10}(\omega_0,\Delta\omega)/\tilde{\text{DOS}}^{\text{vac}}_{N}(\omega_0,\Delta\omega)$ as a function of $\omega_0$ for a gpr of 100 and a $10\times10$ design region with $\frac{\epsilon(\mathbf{r})}{\epsilon_0} \in [1,3.1]$. The quantities $\tilde{\text{DOS}}^{\text{vac}}_{10}(\omega_0,\Delta\omega)$ are the integrals of the product of $\text{DOS}'(\omega)$ and the window function $H_{\omega_0,\Delta\omega,N}(\omega)$, neglecting the pole at $\omega = 0$, when $\epsilon(\mathbf{r}) = \epsilon_0$ throughout the design region, and are also computed for a gpr of 100 and a $10\times10$ design region. The abscissae of all plots are shown for the interval from 0 to 2500 iterations on a linear scale and all ordinate axes are shown for the interval from $10^{-6}$ to 1 on a logarithmic scale.}
\label{fig:varyomegaconv}
\end{figure*}

\begin{figure*}[ht!] 
\centering
\includegraphics[width=0.99\textwidth]{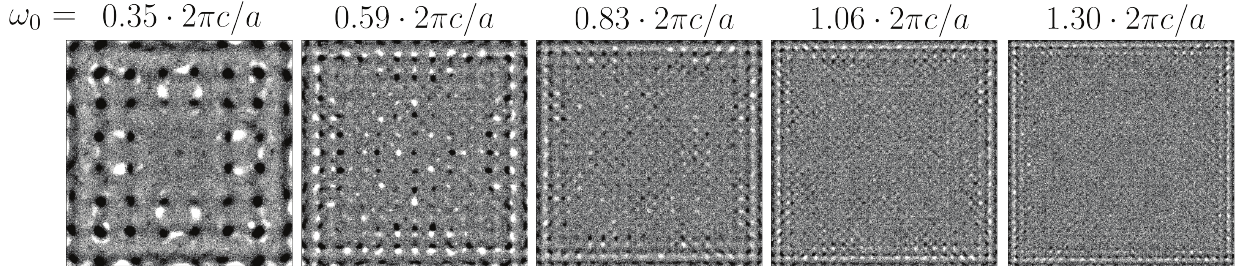}
\caption{Designs optimized for 2500 iterations, a gpr of 100, and a $10\times10$ design region with $N = 10$ as a function of the midgap frequency $\omega_0$. TM polarization was used. Material permittivity values are in the range $\frac{\epsilon(\mathbf{r})}{\epsilon_0} \in [1,8.9]$ using the same color bar as in Fig. \ref{fig:TM_sizes}.}
\label{fig:varyomega8d9}
\end{figure*}

\begin{figure*}[ht!] 
\centering
\includegraphics[width=0.99\textwidth]{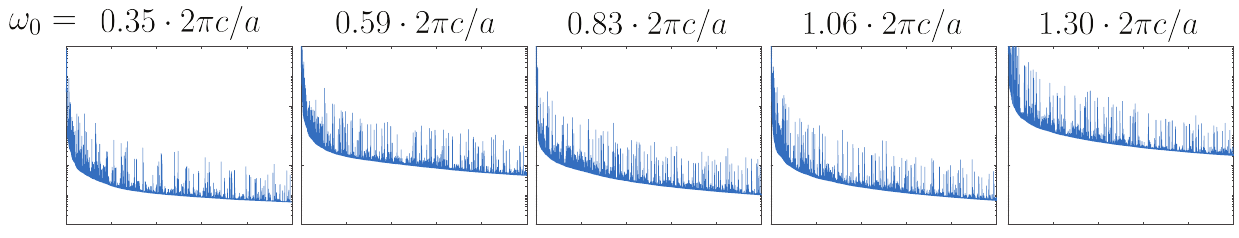}
\caption{Convergence of $\tilde{\text{DOS}}_{10}(\omega_0,\Delta\omega)/\tilde{\text{DOS}}^{\text{vac}}_{N}(\omega_0,\Delta\omega)$ as a function of $\omega_0$ for a gpr of 100 and a $10\times10$ design region with $\frac{\epsilon(\mathbf{r})}{\epsilon_0} \in [1,8.9]$. The quantities $\tilde{\text{DOS}}^{\text{vac}}_{10}(\omega_0,\Delta\omega)$ are the integrals of the product of $\text{DOS}'(\omega)$ and the window function $H_{\omega_0,\Delta\omega,N}(\omega)$, neglecting the pole at $\omega = 0$, when $\epsilon(\mathbf{r}) = \epsilon_0$ throughout the design region, and are also computed for a gpr of 100 and a $10\times10$ design region. The abscissae of all plots are shown for the interval from 0 to 2500 iterations on a linear scale and all ordinate axes are shown for the interval from $10^{-6}$ to 1 on a logarithmic scale.}
\label{fig:varyomega8d9conv}
\end{figure*}

We can compare the results in Fig. \ref{fig:varyomega} and \ref{fig:varyomega8d9} to the result in Fig. \ref{fig:hexagonal}, where we see that the crystal structures for $\frac{\epsilon(\mathbf{r})}{\epsilon_0} \in [1,3.1]$ and for $\frac{\epsilon(\mathbf{r})}{\epsilon_0} \in [1,8.9]$ both easily adapt to a hexagonal supercell. Initializations of grid points with permittivity values were random as described above.  The other parameters used in the optimization were $N = 10$ and the design region was a supercell of a rhombus structure carved out of a $16\times10$ space. In Fig. \ref{fig:hexagonal}, for $\frac{\epsilon(\mathbf{r})}{\epsilon_0} \in [1,3.1]$ $\omega_0 = 1.4\cdot 2\pi c/a$ was used and for $\frac{\epsilon(\mathbf{r})}{\epsilon_0} \in [1,8.9]$ $\omega_0 = 0.83\cdot 2\pi c/a$ was used to ensure that the structures would be on the same scale using the fact that for the homogeneous wave equation $\epsilon(\mathbf{r}) \rightarrow \epsilon(\mathbf{r})/s^2$ results in a trivial rescaling of the band structure $\omega \rightarrow s\omega$~\cite{joannopoulos_photonic_2011}. In both cases, $\Delta\omega = \omega_0/10$ was employed. Using periodic boundary conditions, we transform the hexagonal supercells in Fig. \ref{fig:hexagonal} into a rectangular supercells, then subdivide the rectangular supercells into 10 cells in each dimension (100 cells in total), and then computing the location of the four highest maxima of the absolute value of the Fourier transform that occur at non-trivial positions in each of the cells. We find that the design region generally partitions into subdomains that obey 6-fold symmetry to within an angle of 5$^\circ$. We find no subdomains that obey 4-fold symmetry for $\frac{\epsilon(\mathbf{r})}{\epsilon_0} \in [1,3.1]$, which is to be expected since 4-fold symmetry does not support a gap of $\Delta\omega = \omega_0/10$ for that dielectric contrast~\cite{rechtsman_method_2009}, while we find one subdomain that obeys 4-fold symmetry for $\frac{\epsilon(\mathbf{r})}{\epsilon_0} \in [1,8.9]$, which again abides by earlier results~\cite{rechtsman_method_2009}. 

\begin{figure*}[ht!] 
\centering
\includegraphics[width=0.99\textwidth]{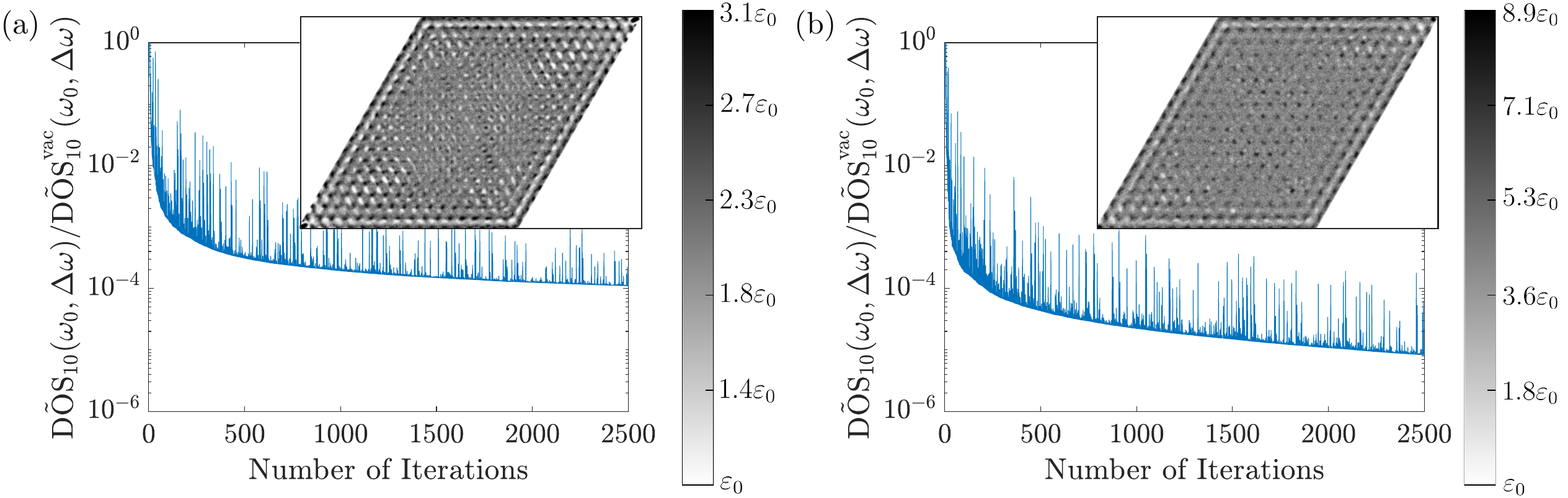}
\caption{Convergence of $\tilde{\text{DOS}}_{10}(\omega_0,\Delta\omega)/\tilde{\text{DOS}}^{\text{vac}}_{N}(\omega_0,\Delta\omega)$ for a gpr of 100 and design regions that were supercells of a rhombus structure carved out of a $16\times10$ space. The quantities $\tilde{\text{DOS}}^{\text{vac}}_{10}(\omega_0,\Delta\omega)$ are the integrals of the product of $\text{DOS}'(\omega)$ and the window function $H_{\omega_0,\Delta\omega,N}(\omega)$, neglecting the pole at $\omega = 0$, when $\epsilon(\mathbf{r}) = \epsilon_0$ throughout the design region, and are also computed for a gpr of 100 and design regions that were supercells of a rhombus structure carved out of a $16\times10$ space. Inset designs were optimized for 2500 iterations in a supercell of a rhombus structure carved out of a $16\times10$ space. TM polarization was used. Material permittivity values are in the range $\frac{\epsilon(\mathbf{r})}{\epsilon_0} \in [1,3.1]$ (a) for which $\omega_0 = 1.4\cdot 2\pi c/a$ was used and $\frac{\epsilon(\mathbf{r})}{\epsilon_0} \in [1,8.9]$ (b) for which $\omega_0 = 0.83\cdot 2\pi c/a$ was used. The color bars indicate the permittivities $\epsilon$ in the various parts of the various design regions.}
\label{fig:hexagonal}
\end{figure*}

We are free to choose any arbitrary design region and our optimization algorithm will adapt the design to the boundary conditions imposed by the particular design region. We show results for various design regions in Figs. \ref{fig:various_designs_eps3d1} and \ref{fig:various_designs_eps8d9} again using random initialization. Ultimately, our approach allows for the discovery of the amorphous or crystalline structure that satisfies a given design region's boundary conditions and we show below that, in the case of polycrystalline structures, we can solve for designs in each unit cell subdomain through a formalism that is analogous to full Brillouin zone integration within the framework of the uniform-source approach. The results in Figs. \ref{fig:various_designs_eps3d1} and \ref{fig:various_designs_eps8d9} are again consistent with the fact that it is easier to open a gap for larger dielectric contrast~\cite{joannopoulos_photonic_2011,rechtsman_method_2009}.

\begin{figure*}[ht!] 
\centering
\includegraphics[width=0.9\textwidth]{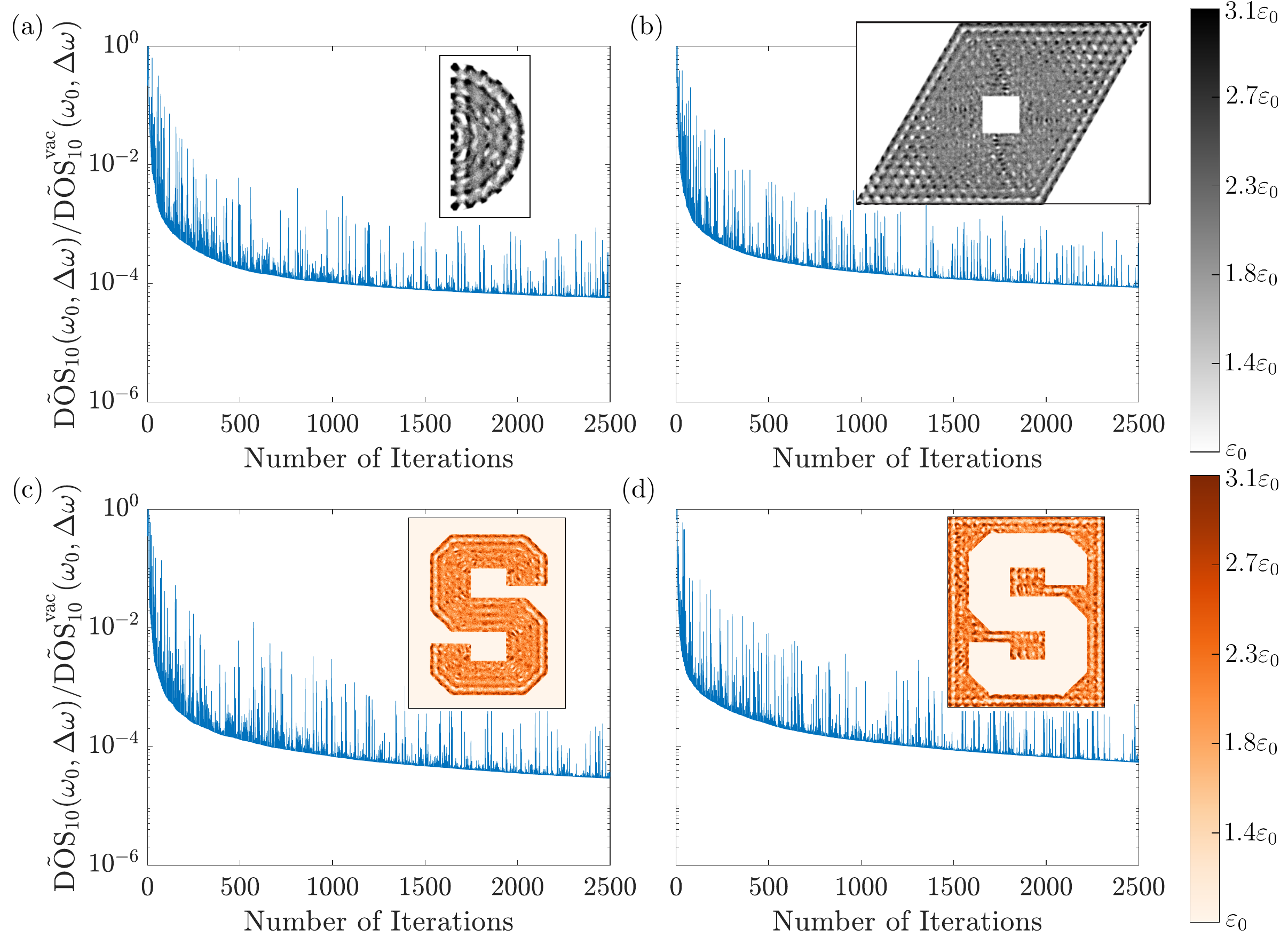}
\caption{Convergence of $\tilde{\text{DOS}}_{10}(\omega_0,\Delta\omega)/\tilde{\text{DOS}}^{\text{vac}}_{N}(\omega_0,\Delta\omega)$ for a gpr of 100 and various design regions. The quantities $\tilde{\text{DOS}}^{\text{vac}}_{10}(\omega_0,\Delta\omega)$ are the integrals of the product of $\text{DOS}'(\omega)$ and the window function $H_{\omega_0,\Delta\omega,N}(\omega)$, neglecting the pole at $\omega = 0$, when $\epsilon(\mathbf{r}) = \epsilon_0$ throughout the design region, and are also computed for a gpr of 100 and the corresponding design region. Inset designs for a half circle (a), a supercell of a rhombus structure with a square cavity (b), a logo of Syracuse University (c), and the complement of the Syracuse University logo (d) were optimized for 2500 iterations for the various design regions. TM polarization was used. Material permittivity values are in the range $\frac{\epsilon(\mathbf{r})}{\epsilon_0} \in [1,3.1]$ and $\omega_0 = 1.4\cdot 2\pi c/a$ was used. All ordinate axes are shown for the interval from $10^{-6}$ to 1 on a logarithmic scale. The color bars indicate the permittivities $\epsilon$ in the various parts of the various design regions.}
\label{fig:various_designs_eps3d1}
\end{figure*}

\begin{figure*}[ht!] 
\centering
\includegraphics[width=0.9\textwidth]{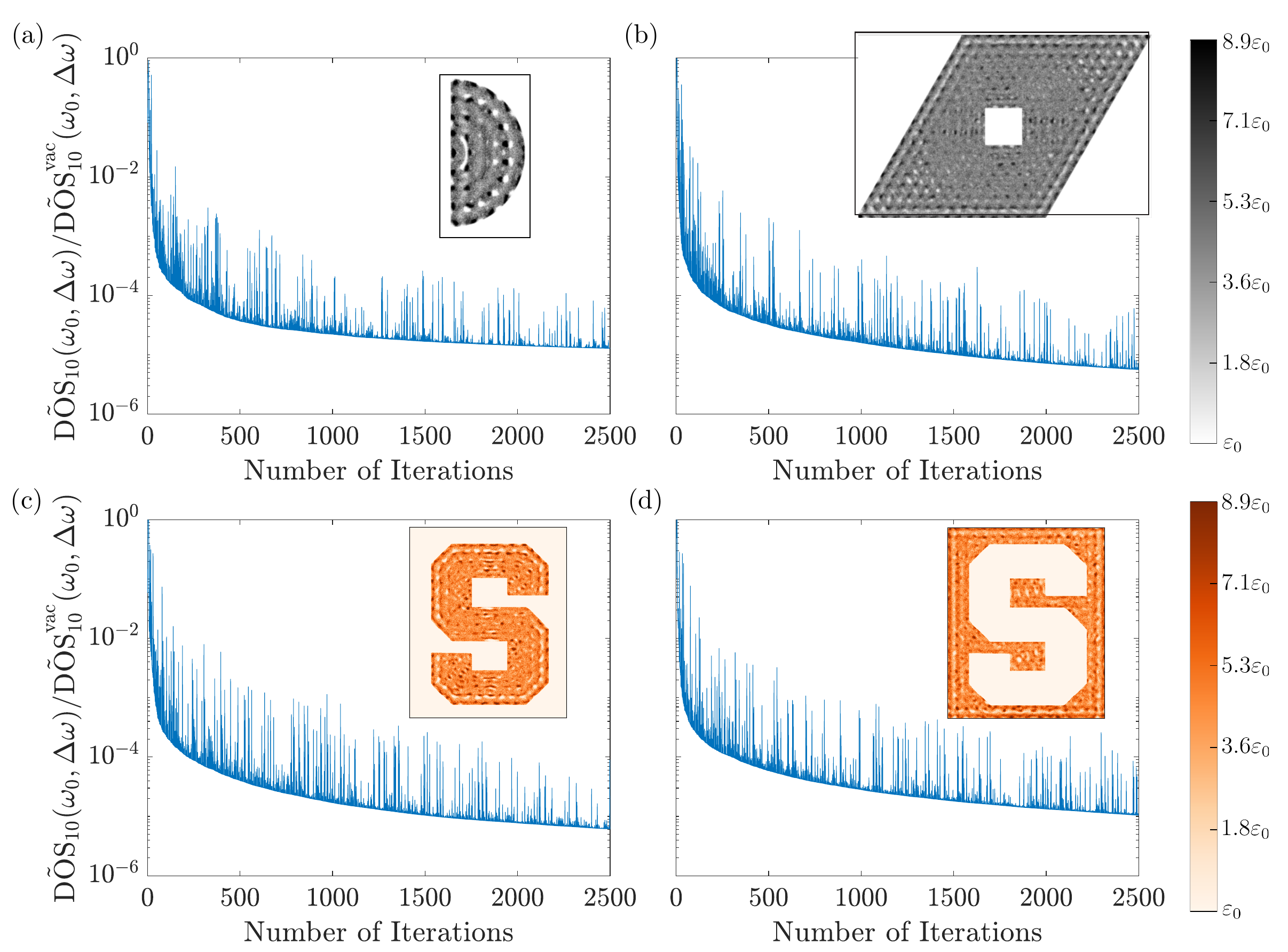}
\caption{Convergence of $\tilde{\text{DOS}}_{10}(\omega_0,\Delta\omega)/\tilde{\text{DOS}}^{\text{vac}}_{N}(\omega_0,\Delta\omega)$ for a gpr of 100 and various design regions. The quantities $\tilde{\text{DOS}}^{\text{vac}}_{10}(\omega_0,\Delta\omega)$ are the integrals of the product of $\text{DOS}'(\omega)$ and the window function $H_{\omega_0,\Delta\omega,N}(\omega)$, neglecting the pole at $\omega = 0$, when $\epsilon(\mathbf{r}) = \epsilon_0$ throughout the design region, and are also computed for a gpr of 100 and the corresponding design region. Inset designs for a half circle (a), a supercell of a rhombus structure with a square cavity (b), a logo of Syracuse University (c), and the complement of the Syracuse University logo (d) were optimized for 2500 iterations for the various design regions. TM polarization was used. Material permittivity values are in the range $\frac{\epsilon(\mathbf{r})}{\epsilon_0} \in [1,8.9]$ and $\omega_0 = 0.83\cdot 2\pi c/a$ was used. All ordinate axes are shown for the interval from $10^{-6}$ to 1 on a logarithmic scale. The color bars indicate the permittivities $\epsilon$ in the various parts of the various design regions.}
\label{fig:various_designs_eps8d9}
\end{figure*}

\subsection{Scaling of computation with grid-point resolution}
We compare the scaling of our approach with a full computation of the photonic density of states using the MPB code~\cite{johnson_block_2001}. The full computation of the photonic density of states scales as gpr$^4$ in 2D, where one factor of gpr$^2$ comes from the area of the design region and the remaining factor of gpr$^2$ comes from evaluation of the contributions to the photonic density of states from all points in reciprocal space in the first Brillouin zone, which we assume to have the same resolution as the real-space grid. Meanwhile, our approach scales as gpr$^2\cdot N\cdot S_x\cdot S_y$, where $S_x$ and $S_y$ are the scale factors by which the supercell is larger than the unit cell in the $x$ and $y$ directions. Thus, when $N\cdot S_x\cdot S_y < \text{gpr}^2$, our approach provides a computational advantage. In Fig. \ref{fig:scaling}, we employed $S_x = 10$, $S_y = 10$, and $N = 10$ and find that for a gpr of 100, the speedup is a factor of 3.3 compared with the expected speedup factor of 10. 

\begin{figure*}[ht!] 
\centering
\includegraphics[width=0.5\textwidth]{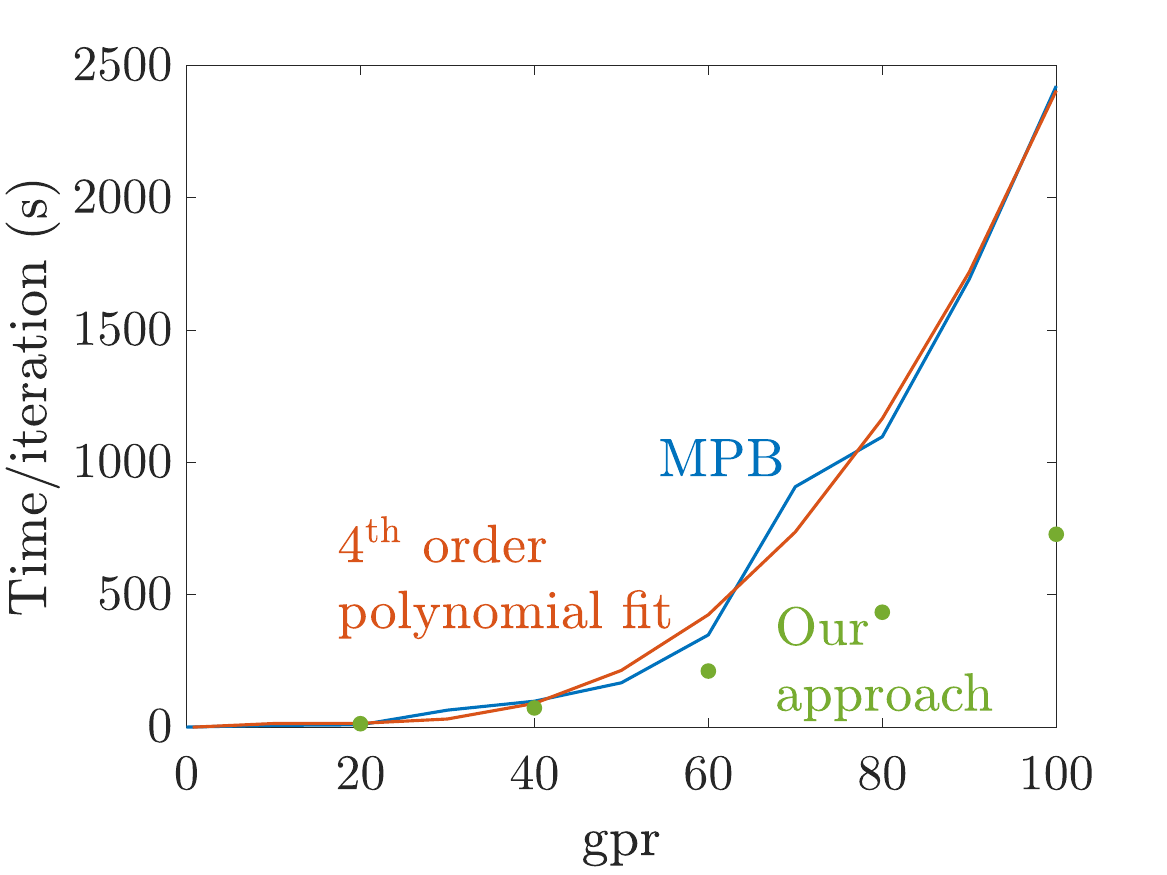}
\caption{Scaling of the computation of $\tilde{\text{DOS}}_{10}(\omega_0,\Delta\omega)/\tilde{\text{DOS}}^{\text{vac}}_{10}(\omega_0,\Delta\omega)$ and a full computation of the photonic density of states using the MPB code~\cite{johnson_block_2001}. The full computation of the photonic density of states scales as gpr$^4$ in 2D, while our approach scales as gpr$^2\cdot N\cdot S_x\cdot S_y$, where $S_x$ and $S_y$ are the scale factors by which the supercell is larger than the unit cell in the $x$ and $y$ directions. For these comparisons, $S_x = 10$, $S_y = 10$, and $N = 10$ were employed and $\epsilon(\mathbf{r})$ was confined to the range $\frac{\epsilon(\mathbf{r})}{\epsilon_0} \in [1,8.9]$ with the random initialization of grid points that has been described in prior sections of the text. A single core was used for all computations.}
\label{fig:scaling}
\end{figure*}

\subsection{Validity of our approach}
We now verify that our algorithm is able to optimize for a target midgap frequency and band gap. We consider a square primitive unit cell and employ the uniform-source approach for various gpr values with $N=10$ in a formulation that is analogous to integration over the full Brillouin zone as described in Section \ref{sec:usrca}. Given the discussion in Section \ref{sec:usrca}, a favorable initialization would be one in which each pixel or voxel is initialized with an extremal permittivity value and the permittivity values are approximately evenly split between the pixels or voxels. We note, however, that we have used the random initialization procedure described in Section \ref{sec:systemsize} to obtain the results below. Once we have found an optimal design, described by $\epsilon(\mathbf{r})$, we find the eigenvalues $\omega_\mathbf{k}^2$ satisfying the generalized eigenvalue problem  
\begin{equation}
\mathbf{\nabla}_\mathbf{k}\times\frac{1}{\mu(\mathbf{r})}\mathbf{\nabla}_\mathbf{k}\times\mathbf{u}_{j,\mathbf{k}}(\mathbf{r}) =  \epsilon(\mathbf{r})\omega_\mathbf{k}^2\mathbf{u}_{j,\mathbf{k}}(\mathbf{r})
\end{equation}
for wavevectors $\mathbf{k}$ along the bandstructure path displayed in Fig. \ref{fig:bandstructures_unitcell} (a). The bandstructure results for gpr values of 10, 20, 30, 40, and 50 are shown in Fig. \ref{fig:bandstructures_unitcell}. We find that for all structures, the midgap frequency is $\omega_0 = 0.40\cdot 2\pi c/a$ in excellent agreement with our target midgap frequency. For the gpr values of 10, 20, 30, 40, and 50, the corresponding band gaps were $\Delta\omega = 0.22\cdot\omega_0, 0.19\cdot\omega_0, 0.18\cdot\omega_0, 0.18\cdot\omega_0,$ and $0.18\cdot\omega_0$, respectively. These bandgap results satisfy our optimization criterion that the band gap be at least as large as $\omega_0/10$. In Fig. \ref{fig:DOSconvergence_unitcell}, we also investigate the convergence of the quantities $\tilde{\text{DOS}}_{10}(\omega_0,\Delta\omega)/\tilde{\text{DOS}}^{\text{vac}}_{10}(\omega_0,\Delta\omega)$ as a function of the number of iterations as the structures that were optimized with the gpr values of 10, 20, 30, 40, and 50 converged toward their optimal designs. We find that the quantities $\tilde{\text{DOS}}_{10}(\omega_0,\Delta\omega)/\tilde{\text{DOS}}^{\text{vac}}_{10}(\omega_0,\Delta\omega)$ are significantly suppressed for all structures.

\begin{figure*}[ht!] 
\centering
\includegraphics[width=0.99\textwidth]{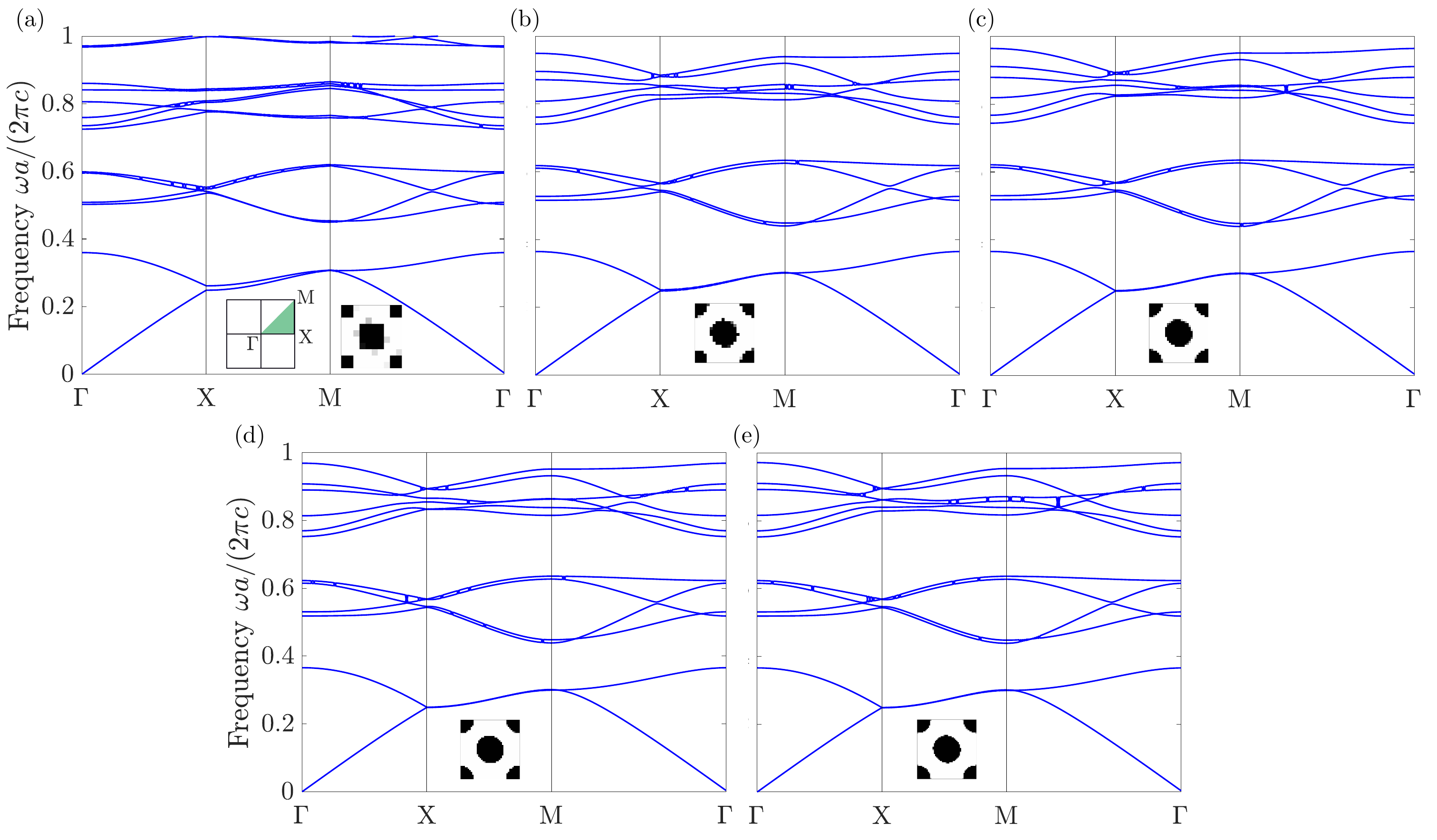}
\caption{The bandstructures obtained for designs optimized with desired values of the midgap frequency and band gap of $\omega_0 = 0.4 \cdot 2\pi c/a$ and $\Delta\omega = \omega_0/10$ for gpr values of 10 (a), 20 (b), 30 (c), 40 (d), and 50 (e) in a square primitive unit cell ($1\times1$ design region) with $N = 10$.  The permittivity values were in the range $\frac{\epsilon(\mathbf{r})}{\epsilon_0} \in [1,8.9]$ with the same color bar as in Fig. \ref{fig:TM_sizes}. The insets of all subfigures contain the optimized designs and the inset of (a) describes the bandstructure path as well. Rotating the inset structure in (a) by 45$^\circ$ reveals similarity to square cross-section rods investigated in the literature~\cite{ashraf_on_2016}.}
\label{fig:bandstructures_unitcell}
\end{figure*}

\begin{figure*}[ht!] 
\centering
\includegraphics[width=0.99\textwidth]{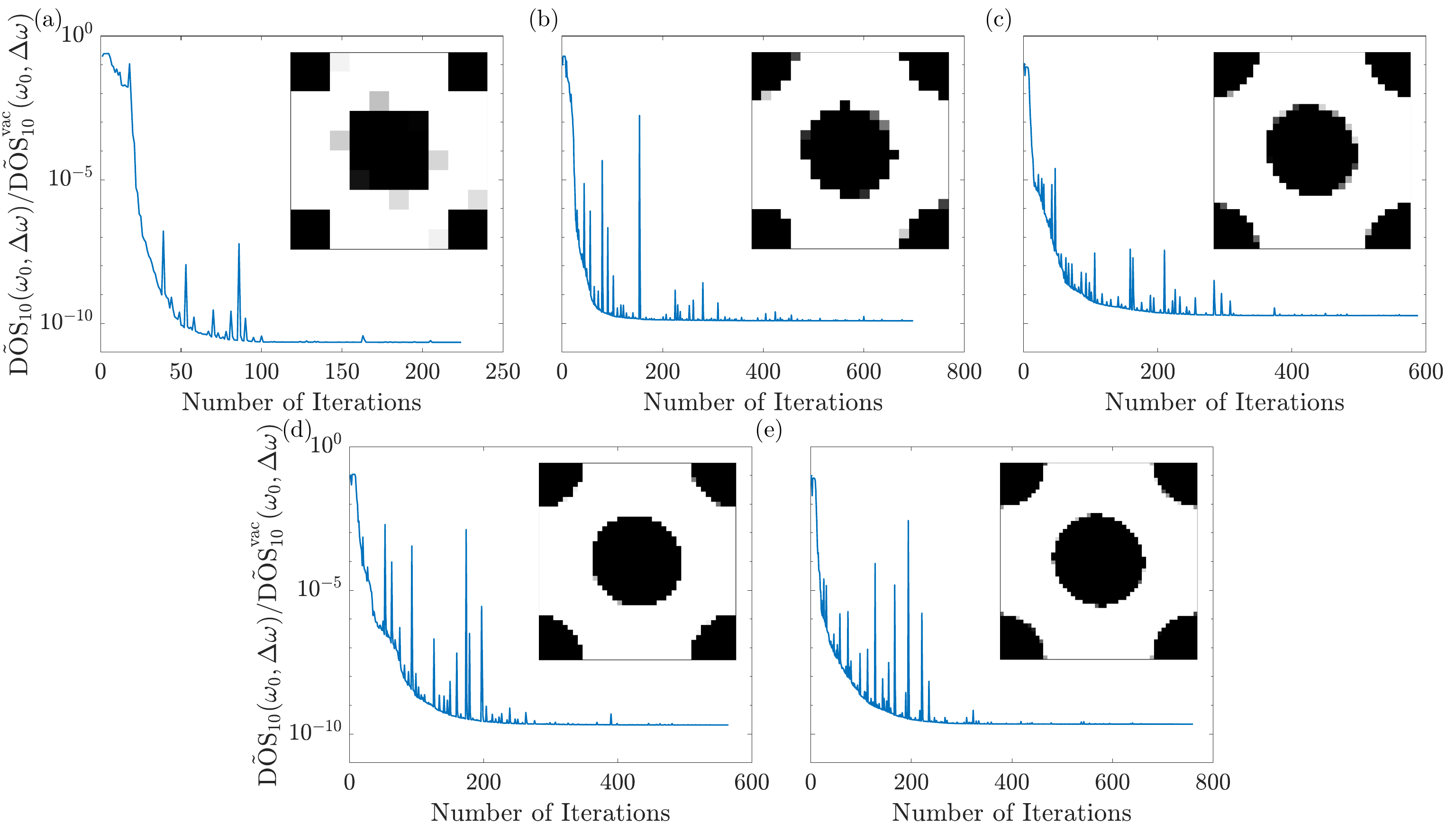}
\caption{We include the convergence of $\tilde{\text{DOS}}_{10}(\omega_0,\Delta\omega)/\tilde{\text{DOS}}^{\text{vac}}_{10}(\omega_0,\Delta\omega)$ as a function of the number of iterations for a gpr values of 10 (a), 20 (b), 30 (c), 40 (d), and 50 (e) in a $1\times1$ design region. The quantity $\tilde{\text{DOS}}^{\text{vac}}_{10}(\omega_0,\Delta\omega)$ is the integral of the product of $\text{DOS}'(\omega)$ and the window function $H_{\omega_0,\Delta\omega,10}(\omega)$, neglecting the pole at $\omega = 0$, when $\epsilon(\mathbf{r}) = \epsilon_0$ throughout the design region, and is also computed for the corresponding gpr values and a $1\times1$ design region. The abscissae of all plots cover the interval from 0 iterations to the total number of iterations for complete convergence on a linear scale and all ordinate axes are shown for the interval from $10^{-11}$ to 1 on a logarithmic scale. The insets of the subfigures contain the optimized designs from Fig. \ref{fig:bandstructures_unitcell}.}
\label{fig:DOSconvergence_unitcell}
\end{figure*}

\section{Conclusion} \label{sec:conc}
Overall, we have presented an approach to optimization for photonic band gaps that can optimize for a specific midgap frequency and band gap in a structure-adaptive manner. This approach is based on optimization of the photonic density of states in a formalism employing a uniform source either in a $\Gamma$-point framework or in a framework analogous to integration over a full Brillouin zone. We generalize the uniform-source approach and standard point-dipole source approach to other hyperuniform collections of sources. We have shown that we can capture known photonic-crystal structures for the TM polarization using the uniform-source approach in two dimensions. We leverage the fact that, when the $\Gamma$-point formalism for the uniform-source approach is employed, TM photonic crystals are observed in supercells for a given minimum supercell size and minimum precision to which the band gap $\Delta\omega$ must be sampled. In addition, we have shown that our $\Gamma$-point and full Brillouin zone formalisms for the uniform-source approach inherently encourage binarized designs even in gradient descent. We have also shown that our approach can treat frequency-dependent optical response. Our approach can ultimately help realize designs that are better adapted to various topologies.

\section{Acknowledgements}
R.K.D. acknowledges financial support that made this work possible from the National Academies of Science, Engineering, and Medicine Ford Foundation Postdoctoral Fellowship Program and from the College of Engineering and Computer Science of Syracuse University. The authors also acknowledge that the work reported on in this paper was substantially performed using Zest, the Syracuse University research computing high-performance computing cluster. Finally, we wish to acknowledge fruitful conversations with Benjamin Strekha, Pengning Chao, and Alejandro W. Rodriguez.




\bibliography{refs_PC}
\end{document}